\documentclass[a4paper,12pt]{article}

\usepackage{float}
\usepackage{amsmath}
\usepackage{amssymb}
\usepackage{feynmf}
\usepackage{caption}
\usepackage{subcaption}
\usepackage{adjustbox}
\usepackage{graphicx}
\usepackage{braket}
\usepackage[all]{xy}
\usepackage[english]{babel}
\usepackage{siunitx}
\usepackage{lipsum} 
\usepackage{lineno,hyperref}
\modulolinenumbers[5]
\usepackage[utf8]{inputenc}
\usepackage{cite}
\usepackage[left=2.75cm,right=2.75cm,top=2cm,bottom=3cm]{geometry}
\usepackage{mathtools}

\usepackage[T1]{fontenc}
\usepackage{url}
 \usepackage{slashed}
 
 \usepackage[dvipsnames]{xcolor}
\usepackage[normalem]{ulem}   

\definecolor{nicered}{rgb}{0.62,0.07,0.07}
\definecolor{nicegreen}{rgb}{0.1,0.5,0.1}
\definecolor{red}{rgb}{1.0, 0, 0}
\definecolor{darkblue}{rgb}{.0,.0,.8}
\definecolor{niceblue}{rgb}{0,0,1}
\definecolor{niceviolet}{rgb}{0.5,0,1.0}
\definecolor{blue}{rgb}{0,0,1}
\hypersetup{colorlinks=true,citecolor=nicegreen,linkcolor=nicered,urlcolor=nicered}

\newcommand{\amc}{\texttt{MadGraph5\_aMC}$@$\texttt{NLO} }
\newcommand{\calchep}{\texttt{CalcHEP} }

\def\ga{g_{a \gamma}}
\def\ge{g_{a e}}
\def\gell{g_{a \ell}}

\def\Gae{\Gamma_{e^+ e^-}}
\def\Gagam{\Gamma_{\gamma \gamma}}
\def\Gachi{\Gamma_{\rm{inv}}}
\def\Gavis{\Gamma_{\rm{vis}}}

\def\sa{\sigma_{a \gamma}}
\def\se{\sigma_{a e}}

\def\gx{g_{a \chi}}
\def\ma{m_{a}}
\def\mx{m_{\chi}}
\def\lumx{\mathcal{L}_\chi}
\def\lumnu{\mathcal{L}_\nu}
\def\aem{\alpha_{ \textrm{em}}}

\def\clsnf{{90\% \,\rm{ C.L.}}}

\def\effCStot{\left( \epsilon \sigma \right)_{e^+e^-\rightarrow a\gamma}}

\def\gmtwoe{(g-2)_e}
\def\gmtwomu{(g-2)_\mu}
\def\dae{\Delta a_e}
\def\damu{\Delta a_\mu}

\newcommand*{\ROMA}{Dipartimento di Fisica, Universit\`a di Roma La Sapienza and INFN, Sezione di Roma, I-00185 Rome, Italy}	
\newcommand*{\INFNFR}{Istituto Nazionale di Fisica Nucleare, Laboratori Nazionali di Frascati, C.P. 13, 00044 Frascati, Italy}

\title{\bfseries  Invisible decays of axion-like particles: constraints and prospects}

\author{ Luc Darm\'e,$^{1}$\footnote{ \url{luc.darme@lnf.infn.it}}  ~Federica 
Giacchino,$^{1}$\footnote{\url{fgiacchi@lnf.infn.it}} ~Enrico 
Nardi,$^{1}$\footnote{\url{enrico.nardi@lnf.infn.it}} \\[0.1em] Mauro 
Raggi$^{2}$\footnote{\url{mauro.raggi@roma1.infn.it}}  } 
\vspace{2cm}	
\date{ \small
    $^1$  \textit{\INFNFR} \\
    $^2$  \textit{ \ROMA}    
	}

\begin{document}
   \maketitle
\begin{abstract}
Axion-like particles (ALPs) can provide a portal to new states of a dark sector.  
We study the phenomenology of this  portal when
the ALP mainly decays invisibly, while its interaction with the standard model sector proceeds essentially 
via its coupling to electrons and/or photons.
 We reanalyse existing limits from various collider and beam dump experiments, including in particular ALP production via electron/positron interactions, in addition to the usual 
 production through ALP-photon coupling.
 We further discuss the interplay  between these limits and the intriguing possibility of explaining simultaneously the muon and electron magnetic moment anomalies. 
 Finally, we illustrate the prospects  of 
 ALP searches at the LNF positron fixed-target experiment PADME, and the  
 future reach of an upgraded  experimental setup.
 \end{abstract}


\newpage
\tableofcontents

\section{Introduction}
\label{sec:intro}

The Standard Model of particle physics (SM) has proven to be an extremely accurate description of nature, predicting with high precision practically all of the observables measured experimentally during the last decades. Yet, its shortcomings are equally well-known, for instance the nature of dark matter, the strong CP problem, the neutrino masses or the origin of the SM flavour structure. Over the years, a large number of Beyond Standard Model (BSM) solutions to these issues have been introduced, but so far none has been experimentally validated. While many of these models rely on introducing new particles around the TeV scale, this is neither a requirement nor, in many cases, the simplest possibility. New, light but feebly interacting particles, singlets under the SM gauge group, can often provide  the most straightforward solution.

One of the best motivated possibilities for  such new light particles are 
pseudo-Nambu-Goldstone Bosons (pNGB) $a$, pseudoscalar particles coupled very feebly to ordinary matter and radiation which arise from an explicit breaking of an approximate 
global symmetry in the UV. Lagrangians involving these fields enjoy an approximate 
shift symmetry under which $a\rightarrow a + const.$, which implies that their 
interactions has to proceed via derivative terms. According to the nature of the quasi-exact global symmetry, 
the corresponding pNGB is often denoted by a specific name: familon~\cite{Wilczek:1982rv},  majoron~\cite{Chikashige:1980ui,Gelmini:1980re}, or  axions~\cite{Peccei:1977hh,Peccei:1977ur,Weinberg:1977ma,Wilczek:1977pj}. More generally, light particles presenting the same types of derivative interactions are also motivated by various BSM theoretical scenarios, such as string theory~\cite{Arvanitaki:2009fg,Cicoli:2012sz},  or by extensions to the SM designed to address specific SM issues such  as the strong-CP 
problem~\cite{Hook:2014cda,Fukuda:2015ana} or the hierarchy problem~\cite{Graham:2015cka}.

From a bottom-up perspective,  particles that enjoy the approximate 
shift symmetry mentioned above are often referred to as axion-like particles (ALPs).\footnote{The name `ALP' is inspired by the QCD axion, the difference  
being that ALPs are not required to solve the strong CP problem, in which case the mass and couplings of the ALP to fermions and photons can be freely taken as independent parameters.}  While the traditional mass range for such new particle typically 
extends in the extra-light, sub-eV region, the possibility of much more massive 
ALPs, in the MeV-GeV range, has received in recent years increasing attention (see. e.g.~\cite{Bauer:2017ris,Dolan:2017osp,Bauer:2018uxu,Feng:2018pew,Harland-Lang:2019zur,Aloni:2019ruo,Dobrich:2019dxc,Ertas:2020xcc,Sakaki:2020mqb,Ertas:2020xcc,Brdar:2020dpr,Kirpichnikov:2020tcf}  for recent works on ALPs coupled to photons/leptons). ALPs in this mass range have been 
invoked for example as a possible solution for both the muon and the electron magnetic moment anomalies~\cite{Chang:2000ii,Marciano:2016yhf,Liu:2018xkx,Abu-Ajamieh:2018ciu,Bauer:2019gfk,Cornella:2019uxs,Kirpichnikov:2020lws} and, interestingly, viable 
solutions to these anomalies require both electromagnetic and leptonic ALP interactions. Last but nor least, recent hints of the production of a $17$~MeV boson in nuclear transitions~\cite{Krasznahorkay:2015iga,Krasznahorkay:2017qfd,Krasznahorkay:2019lyl,Firak:2020eil} have also been interpreted in terms of an hypothetical new pseudo-scalar particle~\cite{Ellwanger:2016wfe,Feng:2016ysn,Kirpichnikov:2020tcf}.

In this work we  explore the possibility that an ALP acts as a portal towards 
a \textit{dark sector} containing other invisible and light SM gauge singlets. Searches for such light and feebly interacting new particles can be 
carried out in experiments which trade a large energy scale for an 
increase in statistics and reduction in backgrounds~\cite{Hewett:2012ns,Essig:2013lka}.
These so-called ``high-intensity frontier'' experiments have attracted a strong interest 
in recent years. Indeed, it has been know for a long time~\cite{Batell:2009di} that for 
interactions mediated by non-renormalisable  operators of dimension $4+n$ ($n=1,..$), 
when the energy scales of the dark sector physics is accessible,  
fixed targets experiments may very well surpass colliders in sensitivity. 
Among various well established search strategies, a particularly powerful one is the search for missing energy in  fixed 
target and beam-dump  experiments, as it does not require any particular  assumption  
on the underlying structure of the dark sector. 

In this work we focus on updating the current limits derived from missing energy searches for an ALP-portal scenario, assuming that the ALP decays dominantly invisibly so that most of the existing limits from visible decays searches do not apply. Furthermore, in simplified ALP models 
generally used to put experimental limits, it is often assumed 
that ALPs have only one  type of interaction, for instance a 
coupling to the photons. Here we work in the more general scenario 
of a MeV-scale ALP coupled to both photons and electrons.
Our aim is to  capture the general phenomenological aspects of an ALP with  
independent couplings to electron and photons, 
studying correlations between processes induced by the different interactions, 
with particular attention to possible changes in the existing limits. 
We cover all available missing-energy searches, including high-energy results from the DELPHI~\cite{Abdallah:2008aa} as well as high-intensity frontier experiments as  BaBar~\cite{Lees:2017lec} and NA64~\cite{Banerjee:2020fue}, and we also present  
projections for Belle-II~\cite{Kou:2018nap}. 
As an illustration of the effect of adding an invisible decay channel for standard long-lived ALP beam dump searches, we further study the limit from the E137 experiment~\cite{Bjorken:1988as} as function of the ALP invisible branching ratio. We show that a large suppression of the visible decay rate is required to alleviate the constraints from this class of experiments. Finally, we discuss the prospects for exploring  ALP models of this type at the Positron Annihilation into Dark Matter Experiment (PADME)~\cite{Raggi:2014zpa,Raggi:2015gza} located in the Laboratori Nazionali di Frascati (LNF) in Italy. As PADME is based on a fixed-target positron beam accelerator setup, its search techniques differ significantly from other missing mass experiments. Near future 
upgrades of this experiment (and of the serving beam infrastructure~\cite{Valente:2017hjt}) will be able to probe a parameter space region 
complementary to NA64. Our results also provide for the first time a complete assessment of the status of ALP based solutions to the 
$\gmtwoe$ and $\gmtwomu$ anomalies. Strikingly, we have found that 
ALPs with masses as low as a hundred MeV can still provide a viable solution, which lies 
within range of upcoming Belle-II results.

A minimal model realising the scenario of an ALP coupled to both photons and electrons
will be presented in Section \ref{sec:model}.
Details about the astrophysical limits and ALP production channels at accelerators will be also presented in this Section. Section \ref{sec:accelerators} is dedicated to 
a recast of existing limits from accelerator experiments, to present the details of 
our estimates for the electron and muon magnetic moments, and 
to a  description of the possibilities for ALP searches at PADME. 
In Section \ref{sec:results} we present a comparison between different limits, 
and finally in Section \ref{sec:conclusions} we draw our conclusions.

\section{The axion-like particle portal}
\label{sec:model}

\subsection{ALP effective Lagrangian}

ALPs are naturally produced by the spontaneous breaking of a global symmetry 
at some large new physics scale $\Lambda$. We follow the effective field theory (EFT) approach and focus on the ALP interaction with leptons and photons after electroweak symmetry breaking. The effective Lagrangian thus contains
\begin{equation}
\mathcal{L} \subset \frac{1}{2}(\partial_{\mu}a)(\partial^{\mu}a)-\frac{1}{2} \ma^2 a^2
+\frac{1}{4} \ga aF_{\mu\nu}\tilde{F}^{\mu\nu}+\sum_{l=e,\mu,\tau} \frac{g_{al}}{2}(\partial_{\mu}a)\bar{l}\gamma^{\mu}\gamma^5l \ ,
\label{eq:lagr}
\end{equation}
where   $\tilde{F}^{\mu\nu}=1/2\,\epsilon^{\mu\nu\alpha\beta}F_{\alpha\beta}$
with $\epsilon^{0123}=-1$. 
In the spirit of ALP phenomenology,  we take the mass $\ma$ and the couplings $\ga$ and $g_{al}$  to be 
independent parameters. The ALP interactions 
are described by dimension-5 operators and carry dimension of inverse mass  (GeV$^{-1}$). The last term in Eq.~\eqref{eq:lagr} is often  rewritten as $-i g_{al} m_l\,a\bar{l}\gamma_5l$ after integrating by part and applying  the equations of motion. However, in passing  from the derivative to the pseudoscalar coupling 
an anomalous contribution to the ALP interaction with photons must be added to the second term in Eq.~\eqref{eq:lagr}~\cite{Bauer:2017ris}. In the rest of this work we will assume that the couplings to quarks and gluon are 
either absent, or negligible compared to $\ga$ and $\ge$ (see for instance~\cite{Ertas:2020xcc} for a 
recent summary of the limits on those couplings).\footnote{We also assume that the $N \gamma$ mixed coupling arising from the ALP interaction in an $SU(2)$-preserving basis is negligible, corresponding to coupling the ALP to $B$ and $W$ vector boson with the same strength (see, e.g. \cite{Dolan:2017osp}).} 
As regards the couplings to the heavier $\tau$ and $\mu$ leptons, 
they would yield additional visible decays channels which in our analysis 
are not relevant. However,   
when we will  briefly consider the prospects 
for the muon $g-2$, $g_{a\mu}$  will clearly play a relevant role. 
We  assume the presence of particles of a  dark sector in which the ALP 
predominantly decays, therefore,  besides the interactions in Eq. \eqref{eq:lagr}, 
the ALP couples derivatively to an axial-vector current  involving dark 
sector states $\mathcal{J}_{5, D}^{\mu}$:
\begin{align}
\label{eq:LJ5}
     \mathcal{L} \supset  \frac{\gx}{2} (\partial_{\mu} a) \mathcal{J}_{5, D}^{\mu}\ .
\end{align}
For example, for a dark Dirac fermion $\chi$ one has 
$\mathcal{J}_{5, D}^{\mu} = \bar{\chi}\gamma^{\mu}\gamma^5 \chi \ $.

The decay widths into two-photons or into an electron-positron pair are given by
\begin{eqnarray}
\Gagam &=&\frac{\ga^2 \ma^3}{64 \pi}  \\
\Gae &=&\frac{\ge^2}{8 \pi} m_e^2 \ma \sqrt{1-\frac{4 m_e^2}{\ma^2}}\,,
\label{eq:Gammavis}
\end{eqnarray}
while the invisible decay width into a pair of dark Dirac fermions reads: 
\begin{eqnarray}
\Gachi &=&\frac{\gx^2}{8 \pi} \mx^2 \ma \sqrt{1-\frac{4 \mx^2}{\ma^2}}\,.
\label{eq:Gammainv}
\end{eqnarray}

Due to the $\ma^3$ scaling of the two-photon decay width, it is clear that one typically expects the photonic channel to dominate the visible decays 
at large ALP masses, except in the case of a significant hierarchy between the photon and the electron couplings. Let us define the visible decay rate of the ALP as $\Gavis = \Gagam + \Gae$. The main assumption in this work is 
dominance of the invisible decays, that is
\begin{align}
\Gachi \gg \Gavis = \Gagam  + \Gae\,.
\end{align}

These partial decay rates have important implications for both visible and invisible decays. Indeed if the size of the detector (for example, the distance between the target and the calorimeter at fixed target experiments) is smaller than the decay length, long-lived particles can mimic an invisible decay event. The typical ALP decay length $\ell_a$ is given approximately by:
\begin{equation}
    \ell_a \sim 0.1\,\textrm{m } \times \left( \frac{100 \ \textrm{MeV} }{ m_a} \right) \times \left[\left( \frac{\gx \mx}{ 0.7 \cdot 10^{-6}} \right)^2 +\left( \frac{\ga \ma}{ 2\cdot 10^{-6}} \right)^2 +\left( \frac{\ge m_e}{ 0.7 \cdot 10^{-6}} \right)^2\right]^{-1} \,.
\end{equation}
where we have assumed $\ma \gg \mx, m_e$ and Dirac dark fermions as an example of a possible dark sector.
From the above equation it is clear that invisible decays will dominate as long as:
\begin{align}
    \gx \mx \gg  \ga m_a\,, \ge m_e \ .
\end{align}
Furthermore, detection in beam dump experiments searching for visible ALP decay will be severely restricted as long as the invisible decay length is too short, which could typically  occur in the region of  parameter space where $\gx \mx \gg 10^{-5}$. 

Let us note that in the case of a significant hierarchy between 
$\ga$ and $\ge$, 
the subdominant coupling will receive important one-loop contributions generated by the 
leading one. Given that the experimental limit are typically much stronger for $\ga$, it is interesting to consider the  electron one-loop contribution proportional to $\ge \aem$. Using the derivative ALP-electron interaction from Eq.~\eqref{eq:lagr} we 
have~\cite{Bauer:2017ris}:
\begin{align}
    \delta \ga  = \frac{\aem \ge }{\pi} B_1 \left(\frac{4 m_e^2}{\ma^2} \right) \ ,
\end{align}
where the scalar loop function $B_1$ is given by:
\begin{align}
    B_1 (\tau) = 1- \tau f(\tau)^2 \ , \qquad  f(\tau) \equiv~&  \begin{cases} \arcsin (1/\sqrt{\tau}) \qquad \textrm{, } \ \tau \geq 1  \\  \bigg[ \frac{i}{2} \log \frac{1+\sqrt{1-\tau}}{1-\sqrt{1-\tau}} + \frac{\pi}{2} \bigg] \qquad \textrm{, } \ \tau < 1\,, \end{cases}  
\end{align}
with $B_1(\tau) \to 1$ for $\tau \to 0$ and  $B_1(\tau) \sim 1/(3\tau) $
for $\tau \gg 1$.  
For $\ma \gg  m_e$  one obtains $\delta \ga \sim 0.001  \ge$, and 
one can thus expect the decay into photons to dominate over the 
decays into electrons as long as $\ma \gtrsim 1$ GeV. 
Note  that, if the correction to 
the ALP-photon coupling is computed using instead the  non-derivative  
coupling to the pseudoscalar electron 
density, one would find a result proportional to  $B_1 (\tau) -1$
which  in the limit of large fermion mass $\tau \gg 1$ does not decouple.
However, as was  pointed out in~\cite{Bauer:2017ris}, the surviving $-1$ precisely cancels 
the anomalous contribution that, as  was mentioned below Eq.~\eqref{eq:lagr}, 
 had to be added to the ALP-photon interaction term  when passing from the derivative 
 of the axial-vector current to the pseudoscalar electron density.
 In the rest of this work, we will assume the coupling $\ga$ to be a completely free parameter, keeping in mind that a ratio $\ga / \ge$ smaller than $\sim 0.001$ 
 is not completely natural, and requires a certain level of tuning of 
 the relevant parameters.

\subsection{Astrophysical and cosmological constraints}

\paragraph{Relic density and CMB.}

An ALP decay channel into invisible particles of a dark sector,    
and the corresponding  suppression of the branching ratios for visible decays, 
can relax the limits from laboratory experiments, as well as from the cosmological 
imprint in the Cosmic Microwave Background (CMB) that could arise from late time 
decays (see e.g. the recent study in ref.~\cite{Depta:2020zbh}).
However,  the presence of new, light, and feebly coupled  degrees of freedom 
has other preeminent consequences in cosmology and especially in astrophysics.

If stable, dark sector particles may constitute all or part of the Dark Matter (DM) content of the universe, a possibility has been studied in the last decade by various groups. 
However, it is relatively hard to suppress  the relic density to an acceptable level in vanilla freeze-out scenarios that exploit a higher-dimensional portal  
(early studies include refs.~\cite{Nomura:2008ru,Dolan:2014ska}). 
Moreover, an additional difficulty is that of evading CMB limits on late-time annihilating DM~\cite{Slatyer:2015jla}. Indeed, a DM candidate 
with mass below $\sim 10$ GeV would  still keep annihilating at the time of 
matter-radiation decoupling, and would produce sizeable distortions of the CMB spectral shape~\cite{Aghanim:2018eyx}.\footnote{Indirect detection limits, in particular  from Fermi-LAT~\cite{Ackermann:2015lka}, also constrain the mass region above~$\sim~200\,$MeV.}  

Nevertheless, a number of ways out from these difficulties have been put forth 
throughout the years. One possibility is to add a secluded annihilation channel 
via a renormalizable operator to fix the correct relic density, and then select a type 
of final states that will decay sufficiently fast in order to avoid all Big Bang nucleosynthesis (BBN) and CMB 
constraints~\cite{Pospelov:2007mp,Darme:2018jmx,Darme:2017glc,Krnjaic:2015mbs}).
Another possibility, that was sketched for example in ref.~\cite{Evans:2019jcs},  
is to use the entropy dilution from the decay of heavier relics to suppress the 
large relic density of keV-scale DM  left over by inefficient annihilation 
via the higher-dimension portal. 
In this case a complete cosmological model should also include the dependency 
on the reheating temperature for the production of the heavy relics, which is also  
subject to specific  constraints. 
More exotic cosmological scenarios, such as a late time phase transition, can additionally 
have strong effects on the resulting relic density (see e.g.~\cite{Dimopoulos:1990ai,Cohen:2008nb} and related literature).
With a set of {\it ad hoc} assumptions the  minimal scenario  outlined below Eq.~\eqref{eq:LJ5} with the dimension five ALP portal and only one dark fermion $\chi$
can also be rendered viable.  For instance, Ref.~\cite{Dolan:2017osp} assumed a resonance setup where $2 \mx \lesssim \ma$ boosts the annihilation rate at earlier times, while 
later on at the BBN and CMB epochs the resonant annihilation channel is quenched
because of the lower temperatures. In such case 
the proper relic density can be obtained for $\ga$ couplings as low as $10^{-5}~\rm{GeV}^{-1}$.
For the lowest masses we will consider (a few MeV), ALP and/or dark sector particle could remain 
in thermal equilibrium until the time of neutrinos decoupling and BBN, and this would yield additional constraints, see e.g.~\cite{Millea:2015qra,Depta:2019lbe,Depta:2020wmr}. 
In this work, given that we do not specify the detailed structure of the dark sector, we 
will simply assume that a mechanism exists that allows to escape possible constraints 
from DM overabundance or from other cosmological arguments.

\paragraph{SN1987 limits.}

The duration of the neutrino burst from the supernova  SN1987A provides 
to this date one of the strongest limit on several types of light new particles.
In fact, particles with mass up to several tens of MeV could be efficiently 
produced in the collapsed core of a SN, and if they can freely escape, 
they would increase the cooling rate of the SN  core and thus shorten the duration of 
the neutrino burst. 
However, if the interactions of these new particles with the surrounding medium are 
not sufficiently feeble, they would remain trapped in the core of the proto-neutron star 
and would not contribute to the  cooling~\cite{Jaeckel:2017tud,Raffelt:2006cw}.
According to the recent study in Ref.~\cite{Lucente:2020whw}
this would  for example happen for ALP-photon couplings 
$\ga \gtrsim 10^{-6}-10^{-5}$ while, to the best of our knowledge, 
no analogous limit has been derived for the case of a purely electrophilic ALP 
(this might be due to the fact that ALP-electron couplings are usually considered in 
combination with ALP-quark interactions, which are expected to dominate because of 
the induced couplings of the ALP to the nucleons). The above result for $\ga$ implies that 
for practically the whole parameter space relevant for this work, the ALP will remain trapped 
and thus SN1987A does not provide useful constraints. 
Only in the lowest part of  parameter space ($\ga \lesssim 10^{-5} ~\rm{GeV}^{-1}$)
trapping via ALP-photon interaction might be avoided. 
However, due to the intrinsic difficulty in pinning down the precise boundary between 
the trapping and free streaming regimes, it is  safer to 
assume that also this region remains at least marginally compatible 
with the SN bound. 
The limits on the ALP couplings to SM particles, however, are of little 
importance in our case, since the  decay/inverse-decay 
process $a \leftrightarrow \chi \bar{\chi}$ changes 
dramatically the discussion. 
In fact,  as was noted in~\cite{Dolan:2017osp}, even when the ALP is trapped because 
of  sizeable $\ga$ or $\ge$  couplings, the dark fermions $\chi$ might 
still escape and drain energy from the SN core, because  $\chi$-trapping  is much more 
difficult to achieve due to the dimension-5 nature of the ALP portal. 
However, Ref.~\cite{Dolan:2017osp} only considered the scattering processes 
$\bar\chi \chi\leftrightarrow \mathrm{SM\, SM}$ for which the rates are heavily suppressed.
We find instead that in the regime in which the ALP is trapped, $\chi$ interactions 
with on-shell ALPs 
give the dominant contribution to reduce the  $\chi$ mean-free-path 
and to keep them trapped. 

The SN limits for light dark matter  are typically derived in two-steps. 
First, one estimates the luminosity $\lumx$ of the black body emission of 
fermions from a $\chi$-sphere of radius $R_\chi$ defined, in analogy to the 
usual neutrino-sphere, as the boundary surface  between the trapping and 
free streaming regimes. 
For a Dirac fermion $\chi$ we have (see e.g.~\cite{Chang:2018rso}):
\begin{align}
        \lumx = \frac{g_\chi}{2 \pi} \int_{\mx}^\infty dE \frac{ E (E^2 - m^2)}{e^{E/T(R_\chi)}+1} \xrightarrow[]{\mx \ll T} \frac{7 g \pi^3}{240} R_\chi^2\, T(R_\chi)^4\,,
\end{align}
where $T$ is the temperature and $g$ the number of degrees of freedom
of the emitted particle ($g = 4$ for a Dirac fermion). Using the core profile from~\cite{DeRocco:2019jti} 
and the neutrino luminosity $\lumnu \sim 3\cdot 10^{52} \textrm{ erg \ s}^{-1}$~\cite{DeRocco:2019jti,Raffelt:2006cw}  we obtain that the typical radius for which 
$\lumx \lesssim \lumnu$ is $R_\chi \sim 20\,$km for $\mx$ around $10\,$MeV. 
The second step is that of  estimating $R_\chi$ which, as anticipated, is   
mainly determined by the inverse-decay process $ \chi \bar \chi \to a$,  
as function of  $g_{a\chi}$.
Using the narrow width approximation ($\Gamma_a \ll m_a$) which is always 
very accurate for the weakly interacting $\chi$, the relevant cross-section reads:
\begin{align}
\label{eq:resCS}
    \sigma_{res} = \frac{\pi \gx^2 \mx^2}{2\sqrt{1-4 \mx^2/\ma^2}} \delta (s - \ma^2) \equiv \tilde{\sigma}_{res}  \delta (s - \ma^2) \ .
\end{align}
 The number of interactions a $\chi$ particle suffers 
 along an outward trajectory from $R_\chi$ is:
\begin{align}
\label{eq:navSN}
N_{int} = \int_{R_\chi}^{R_{far}} dr \left[ \frac{g_\chi}{n_\chi (T_{R_\chi})} \int d^3\Pi_\chi d^3\Pi_{\bar{\chi}} f_{\chi}(T_{R_\chi})f_{\bar{\chi}}(T) 4E_\chi E_{\bar{\chi}} \sigma_{res} \right] \ ,
\end{align}
where $d^3\Pi$ is the standard Lorentz-invariant phase-space, $R_{\chi}$ is the radius of the dark sector sphere determined above,  $f_\chi = (\exp(m_\chi/T)+1)^{-1}$ is the Fermi-Dirac 
statistical distribution, and $R_{far} \sim 100\,$km corresponds to the outer layers of the proto-neutron star, out of which the profiles are not accurate. 
Altogether, Eq.~\eqref{eq:navSN} is simply the thermal average (taken at $T_{R_\chi}$) of the number of interactions that a fermion $\chi$ emitted from $R_\chi$ has with the surrounding 
halo of $\bar \chi$'s to resonantly produce an ALP.
Given the simple form of $\sigma_{res}$ and the symmetric structure in $\bar{\chi}$ and $\chi$, 
the  integral can be evaluated straightforwardly.
Using standard integration techniques from the dark matter playbook 
(see e.g. the nice Appendices in~\cite{DEramo:2017ecx}) we have
\begin{align}
    d^3\Pi_\chi d^3\Pi_{\bar{\chi}}  = \frac{|p_\chi||p_{\bar{\chi}}|}{32\pi^4} dE_\chi dE_{\bar{\chi}} d \cos \theta\,, 
\end{align} 
where $\theta$ is the angle between $\chi$ and $\bar{\chi}$,  and $|p_\chi|, |p_{\bar{\chi}}|$ their three-momenta. Using also  
\begin{align}
\label{eq:sdev}
s = 2 m_\chi + 2 ( E_\chi E_{\bar{\chi}} - |p_\chi||p_{\bar{\chi}}| \cos \theta )\ ,
\end{align}
and absorbing the angular integration by means of the delta function in Eq.~\eqref{eq:resCS}, we obtain the simple expression:
\begin{align}
N_{int} = \int_{R_\chi}^{R_{far}} dr \left[ \frac{g_\chi}{n_\chi}  \int_{\mx}^\infty \int_{\mx}^\infty dE_\chi dE_{\bar{\chi}} \left(\frac{E_\chi E_{\bar{\chi}}}{(e^{E/T(r)}+1)(e^{E/T(R_\chi)}+1)} \Theta (E_1,E_2)\right) \frac{\tilde{\sigma}_{res}}{16 \pi^4} \right] \ , 
\end{align}
where  $\Theta(E_{\chi},E_{\bar{\chi}}) =1$ in the 
resonance region defined by $s = m_a^2$, and $0$ outside. Requiring at least one 
inverse-annihilation $N_{int} \geq 1$, and taking  as an example $\mx = \ma / 3 $, 
leads to the trapping limit on the $\chi$-ALP coupling  
\begin{align}
\label{eq:SNlimit}
\gx \mx  \gtrsim 10^{-6} \ ,
\end{align}
that holds for $\mx $ between $\sim 5 -  50$ MeV.\footnote{An alternative approach is that of estimating directly the mean free path for a particle $\chi$ of energy $E_\chi \gtrsim m_a/2$ and of mass in the tens of MeV range, emitted from the $\chi$-sphere $R_\chi$, and 
to require a mean free path of the order of $1\,$km. Following this procedure we have found 
a value of $\gx \mx$ of the same order of magnitude than in Eq.~\eqref{eq:SNlimit}.}
Although Eq.~\eqref{eq:SNlimit} is just an order of magnitude estimate, that moreover has been derived for the specific case of a DM Dirac fermion, it clearly shows that trapping can be achieve with reasonable dark sector couplings  (typically of the order of the experimental limit 
on $\ge$, which is consistent with our initial assumption   of ALP decays dominated by 
invisible channels). We can then conclude that in our case 
the ALP limits from SN1987A cannot be applied.

Altogether, we conclude that the limits from SN1987 cooling can be escaped by trapping the light dark matter in the proto-neutron star via its resonant annihilation into ALP. In the following we will thus simply assume that the dark coupling $\gx$ is large enough to trap $\chi$ and will focus instead on the collider signatures directly. Note that since the dark matter still does travel before annihilating, one should more rigorously consider the energy flux mediated by this process, using quantities like the Rosseland mean opacity~\cite{Raffelt:2006cw}. In that case, the precise structure of the dark sector should be fleshed out, in particular if it contains more than one field. Nonetheless, Eq.~\eqref{eq:SNlimit} is several orders of magnitude below what will be typically required to maintain an invisible ALP while having $\ge, \ga$ around the state of the art limits from laboratory based experiments, so we still expect the later constraint to the dominate.

As a final comment, recent works~\cite{Sung:2019xie,Lucente:2020whw} have further pointed to the possible importance of the dark sector particles in the description of the actual supernova explosion. Deriving actual constraints from such considerations would however require an actual simulation of the supernova including the presence of a dark sector.

\subsection{ALPs production}
\label{sec:totxsecLO}

\begin{figure}[h]
\begin{center}
\begin{tabular}{cc}
\begin{subfigure}[b]{0.3\textwidth}
\includegraphics[width=\textwidth]{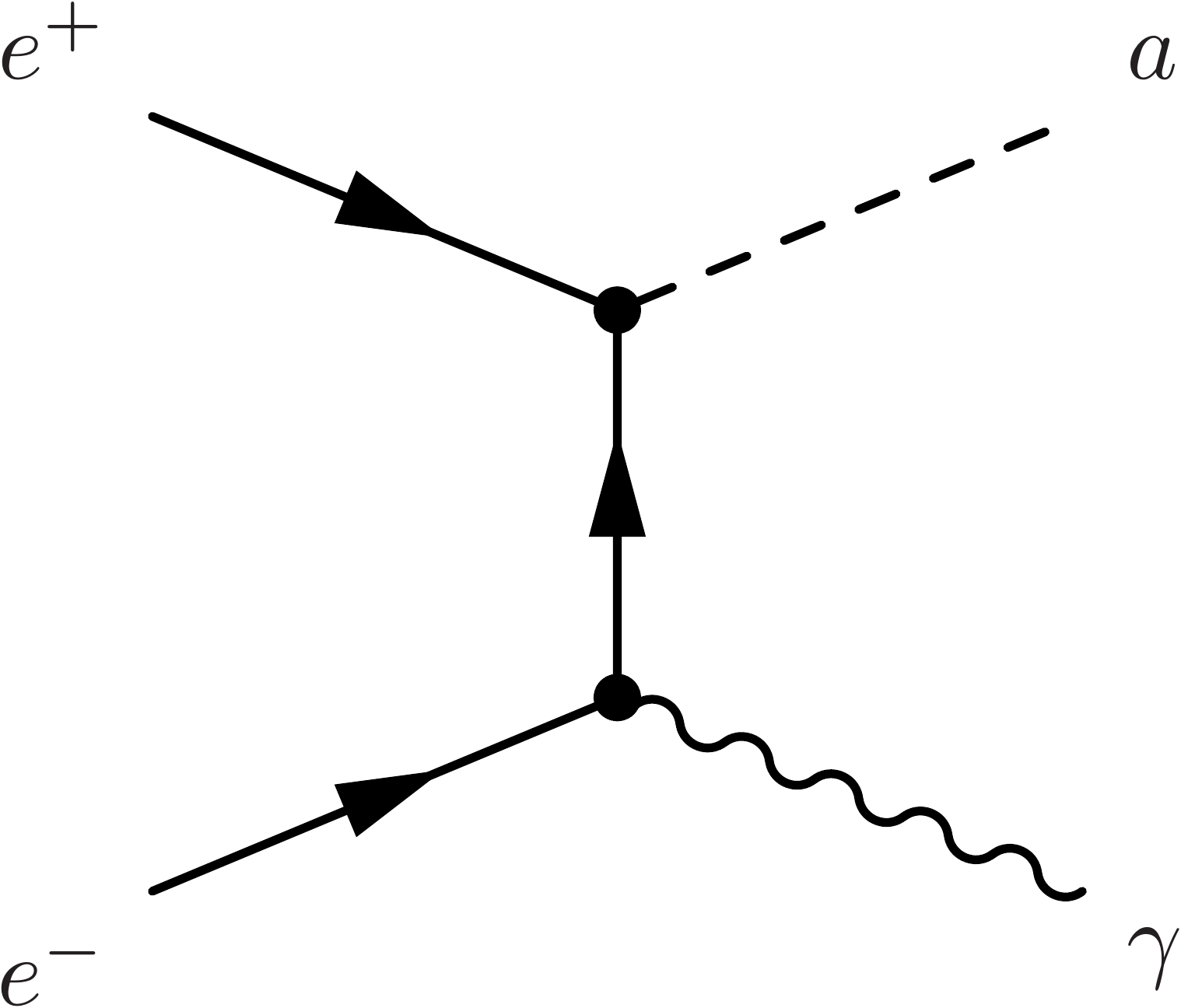}
\caption{}
\label{t-ann}
\end{subfigure} &\begin{subfigure}[b]{0.3\textwidth}
\includegraphics[width=\textwidth]{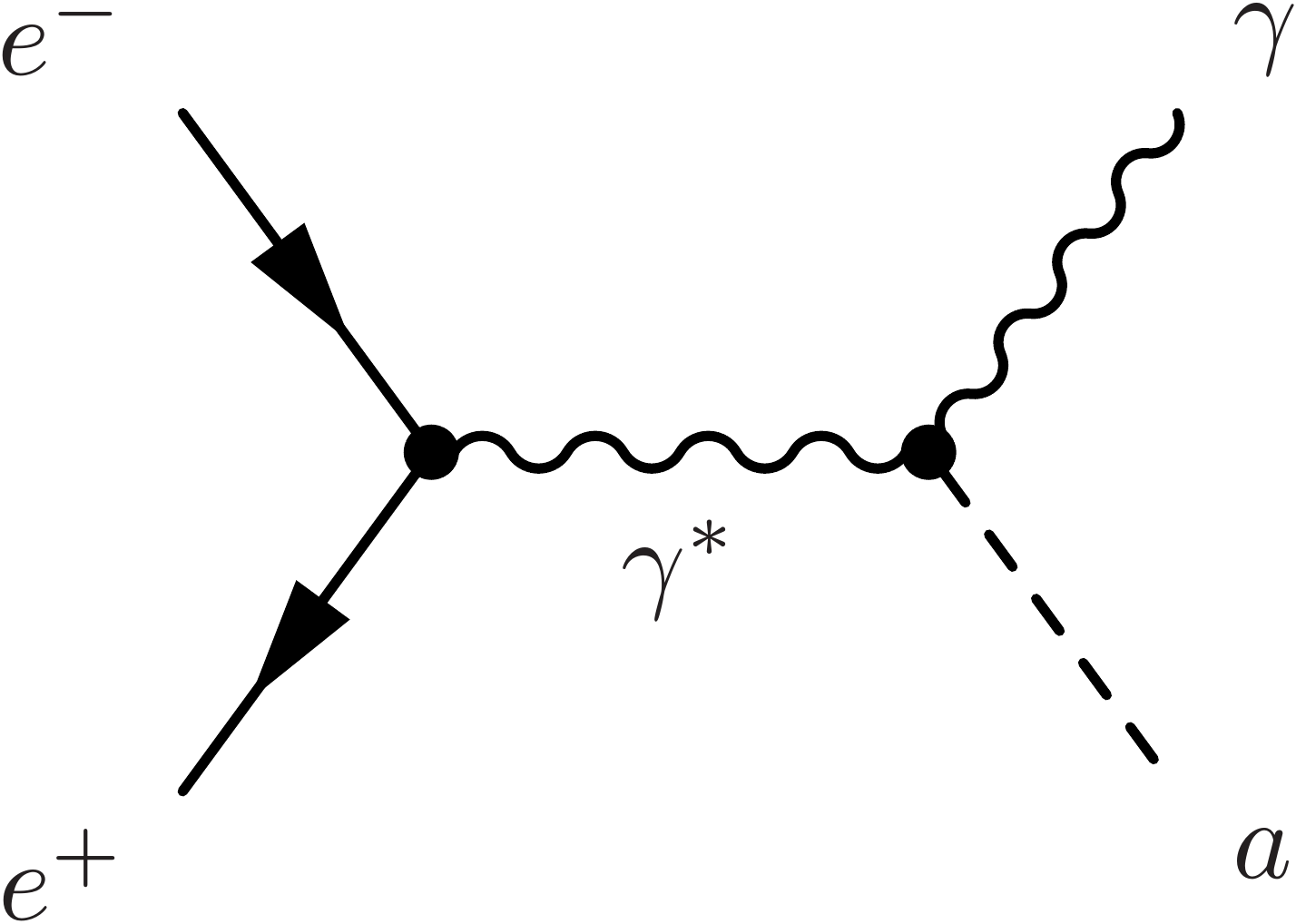}
\caption{}
\label{s-ann}
\end{subfigure}\\
\begin{subfigure}[b]{0.3\textwidth}
\includegraphics[width=\textwidth]{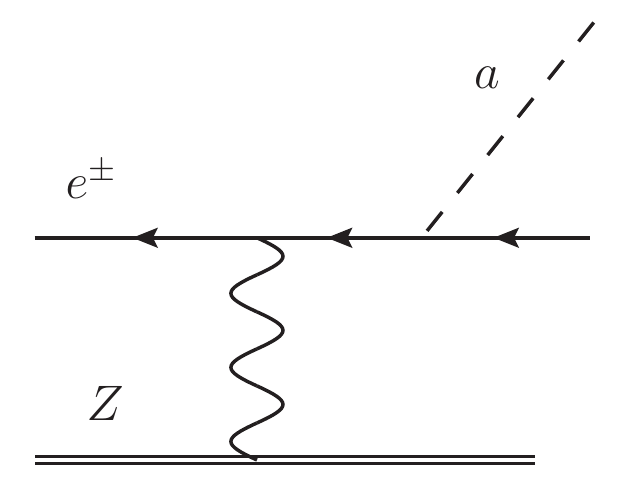}
\caption{}
\label{fig:Brem}
\end{subfigure} &\begin{subfigure}[b]{0.3\textwidth}
\includegraphics[width=\textwidth]{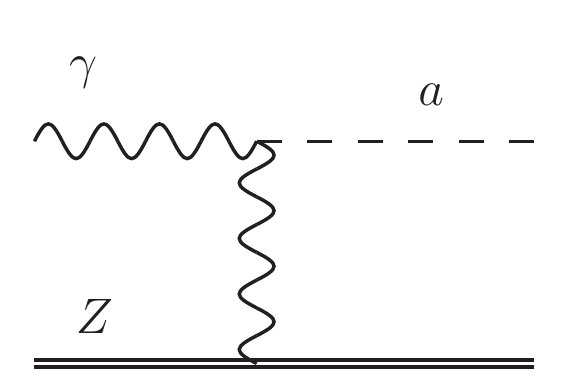}
\caption{}
\label{fig:primakoff}
\end{subfigure}
\end{tabular}
\end{center}
\caption{Feynman diagrams for ALP production: (\subref{t-ann}) $t$-channel annihilation process $e^+e^-\rightarrow \gamma + a$; (\subref{s-ann}) $s$-channel annihilation process $e^+e^-\rightarrow \gamma + a$; (\subref{fig:Brem}) Bremsstrahlung production off a target nucleus $N$ of atomic number $Z$,  $e^\pm  \rightarrow N + e^\pm  + a $; (\subref{fig:primakoff}) Primakoff production from a secondary photon $\gamma + N \rightarrow N + a$.}
\label{diagrams}
\end{figure}

ALPs can be produced through many processes in lepton-based experiments. Since we focus mostly on missing energy searches, we will typically require an additional photon final state to trigger on the events. The main production channels in lepton beam accelerators 
are shown in Fig.~\ref{diagrams}. For positron beam experiments or $e^+ e^-$ colliders, the dominant processes are  annihilation in the $t$ or $s$-channel $e^+ + e^- \rightarrow \gamma + a$ (Fig.~\ref{t-ann} and~\ref{s-ann}), the former depending on the electron coupling $\ge$ and the latter on the ALP-photon coupling $\ga$. 
In electron-beam experiments, the dominant processes are either ALP direct bremsstrahlung production on a target nucleus $N$ via the $\ge$ coupling shown in Fig.~\ref{fig:Brem}, or conversion of bremsstrahlung photons from  
$e^- N \to e^- N \gamma$ into ALPs via the process  $\gamma N \to  N a$ 
shown in Fig.~\ref{fig:primakoff} that depends on $\ga$.
Several other ALP production channels  have been considered in the literature. However, they are either subdominant in the experimental setups we are  considering here, or 
not  relevant because they do not provide the final states that have been 
searched for by the experimental collaborations.
For instance, the `photon-fusion' process $e^+ e^- \to e^+ e^- (\gamma^* \gamma^*) \to e^+ e^- a$, which can have significant rates in $e^+ e^-$ colliders~\cite{Marciano:2016yhf,Bauer:2017ris} does not lead to a mono-photon 
final state
(unless a $\gamma$ is radiated from a fermion line, in which case the process
is of higher order), and is subdominant with respect to  Primakoff production  
in  experiments such as NA64  in which the beam impinges onto a high-$Z$ dump. 
ALPs can be also produced via the inverse-Compton  process $\gamma e^- \to a e^-$ 
that depends on $\ge$. This could be an effective production channel 
in beam dumps where a large population of secondary photon is generally  created~\cite{Brodsky:1986mi,Dent:2019ueq}. However,  we have checked that 
inverse-Compton  remains a sub-dominant production process with respect 
to ALP bremsstrahlung from the primary electrons of 
the beam.\footnote{
A special case is positron annihilation on atomic electrons via the diagram in 
Fig.~\ref{t-ann} when $2 m_e E_{e^+} \approx \ma^2$
and the emitted photon is soft.
However, this process can be seen as a radiative correction to the resonant 
annihilation $e^+ e^- \to a$ which then should be also included.  
The importance of resonant annihilation for positron beam fixed target 
experiments was  first pointed out in Ref~\cite{Nardi:2018cxi} in relation 
to dark photon searches at PADME,  and was later included in a reanalysis 
of electron beam experiments in which the positrons arise as secondaries~\cite{Marsicano:2018glj,Marsicano:2018krp}.
At NA64, given the strong energy cut of $50$ GeV, resonant annihilation 
would typically contribute for $\ma \sim 200 -300$ MeV,  and could possibly help 
the experiment to overcome BaBar constraints in this narrow range. 
A dedicated study of this possibility,  similarly to what has been done for instance for E137~\cite{Marsicano:2018krp}, could indeed be interesting, and is left    
for future work.}

\paragraph{Positron-electron annihilation.}

We first focus on the associated annihilation processes, which is 
relevant for positron beam and $e^+ e^-$ experiments. 
The total annihilation cross-section in the laboratory frame is given by:
\begin{align}
\sigma=\ &\sa+\se+\sigma_{int}=
\aem\, \ga^2 \frac{(s+2m_e^2)(s-\ma^2)^3}{24\beta s^4} \nonumber\\
&+ \aem\, \ge^2 m_e^2\frac{-2\ma^2\beta s+(s^2+\ma^4-4\ma^2m_e^2) \log \frac{1+\beta}{1-\beta}}{2(s-\ma^2)s^2\beta^2} \nonumber\\
&+ \aem\, \ga\,\ge m_e^2 \frac{(s-\ma^2)^2}{2\beta^2s^3} \log \frac{1+\beta}{1-\beta}\,,
\label{eq:totxsec}
\end{align}
where $\aem=e^2/(4\pi)$ is the  electromagnetic fine structure constant, $m_e$ the electron mass, $\beta=\sqrt{1-\frac{4m_e^2}{s}}$ and $\sqrt{s}$ is the center-of-mass energy. In
the first equality 
$\sa$, $\se$ and  $\sigma_{int}$ refer respectively to the contributions 
to the annihilation from   
$s$-channel photon exchange   (Fig.~\ref{s-ann}), 
$t$- and $u$-channel electron exchange (Fig.~\ref{t-ann}), 
 and  to their interference. 
\begin{figure}[t]
\begin{center}
        {\includegraphics[height=0.52\linewidth]{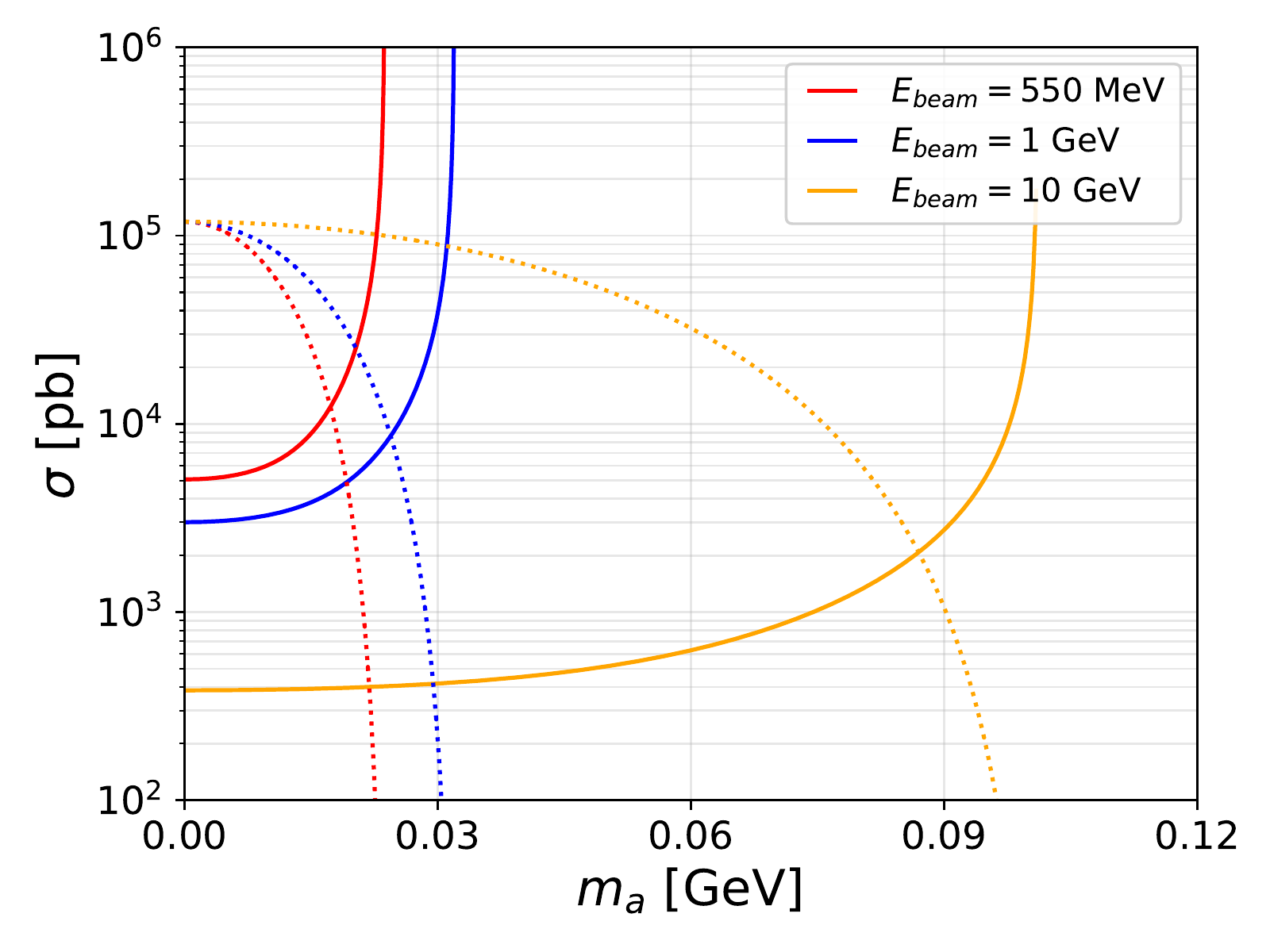}}
        \caption{Contributions to the $e^+e^-$ annihilation cross-section as a function of the ALP mass $\ma$ for three different values of the beam energy $E_{beam}=0.55,1,10 \text{ GeV}$.
        Continuous lines refer to $t$- and $u$-channel electron exchange 
                 (Fig.~\ref{t-ann}) with $\ge=1\, \rm{GeV}^{-1}$. Dotted lines 
        to $s$-channel photon exchange        (Fig.~\ref{s-ann})
        with $\ga=1\, \rm{GeV}^{-1}$.}
\label{fig:xsecEbeam}
\end{center}
\end{figure}
The plots in Fig.~\ref{fig:xsecEbeam} show the different behaviours of $\sa$ and $\se$ as a function of the ALP mass $\ma$ for various beam energies. Interestingly, the two processes have a starkly different behaviour, due to the presence of a resonant enhancement in the soft-photon limit in the $t$- and $u$-channel processes of Fig.~\ref{t-ann}. On the 
other hand, the electron and photon exchange contributions are  roughly of the same 
order when the two couplings  $\ga$ and $\ge$ satisfy 
\begin{align}
\label{eq:condition}
    \ga \approx \ge \frac{m_e}{\sqrt{s}} \ .
\end{align}
This relation underlies much of the phenomenology of the next sections: photon-mediated ALP production is relevant at large center-of-mass energies, while the converse 
is true for electron-mediated processes. In particular, the associated ALP-$\gamma$ production  is characterised by a resonant enhancements when $\sqrt{s} \sim \ma$~\cite{Nardi:2018cxi}. This implies  that high-statistics low-energy positron-beam experiments can be 
particularly efficient in constraining the ALP-electron coupling since they can 
exploit the resonant process to enhance dramatically the production of ALPs.
As regards the interference term in Eq.~\eqref{eq:totxsec}, it presents the 
downsides of both the pure electron and pure photon contributions, and is 
subdominant in all the parameter space we have probed. Indeed, considering the ``maximum interference'' scenario in which  $\ga$ and $\ge$ fulfill the condition in Eq.~\eqref{eq:condition}
we see that $\sigma_{int}$ does not get resonantly enhanced at $\sqrt{s}\sim \ma$, 
and that in the large $ \sqrt{s} \gg \ma$ limit it  scales with an additional suppression factor of $m_e/\sqrt{s}$.

\paragraph{Bremsstrahlung production.}

Let us now consider the most relevant production channels  for electron 
beam experiments. 
In this case, ALP production relies on the electron or photon 
interaction with the electromagnetic field sourced by an atomic 
nucleus $N$.
In the first case the electron interacts with the nucleus via  a 
virtual photon, and recoils emitting an ALP  via  bremsstrahlung. 
The physics underlying this process has been known for several 
decades (see~\cite{Tsai:1973py} for a review of early studies).
The photon interaction with the nucleus is described in terms of a 
form factor with one elastic and one inelastic component: 
$G_2(t) = G_2^{el} + G_2^{in} $. Denoting with 
$t$ the virtuality (i.e. the squared momentum) of the 
virtual photon,  
we have:\footnote{Note that our expression for $G_2^{in}$ differs 
from the one given in Eq.(A19) of Ref.~\cite{Bjorken:2009mm} 
due to a spurious square for the term  in  the second parenthesis 
appearing in that reference,  see also Ref.~\cite{Jodlowski:2019ycu}.} 
\begin{align}\nonumber
	G_2^{el}&=\left(\frac{a^2 t}{1 + a^2 t}\right)^2 \left(\frac{1}{1 +
	t/d}\right)^2 Z^2\,, \\
	G_2^{in}&=\left(\frac{{a^\prime}^2 t}{1+{a^\prime}^2 t}\right)^2 \left(\frac{1 +
		\frac{t}{4m_p^2}(\mu_p^2 - 1)}{\left(1 +
		\frac{t}{0.71\,\mathrm{GeV}^2}\right)^4}\right) Z \, ,
\end{align}
where $Z$ is the nucleus atomic number, $m_p = 0.938\,$GeV 
is the proton mass, $\mu_p = 2.79$ is the nuclear magnetic 
dipole moment. The parameters 
 $a$~\cite{Schiff:1951zza} and $a'$~\cite{Tsai:1973py} 
are related to the atomic size in the elastic and inelastic case, and are determined 
such that in the limit of complete screening one reproduces the numerical results, 
while $d$ parametrises the inverse 
nucleon square radius. Their values 
are~\cite{Schiff:1951zza,Tsai:1973py}
\begin{align}
\label{eq:defad}
 a = 111 \frac{1}{m_e Z^{1/3}} ,\ \  a' = 773 \frac{1}{m_e Z^{2/3}}, \ \ 
 d = 0.164 ~\textrm{ GeV}^2 A^{-2/3} \, .
\end{align} 
As noted in ref.~\cite{Celentano:2020vtu}, the exchange of a very soft or very hard 
photon is suppressed due to either the screening from the electrons in the atomic 
cloud when $a^2 t, a'^2 t \ll 1 $ or from the finite nuclear size in the opposite limit
 $ t \gg d$. 

Although both the $\ga$-driven Primakoff and $\ge$-driven bremsstrahlung processes 
share the same effective interaction with the target nuclei, 
the former is strongly enhanced with respect to the latter. Indeed, in 
the regime $m_e \ll \ma \ll E_0$ (with $E_0$ the incoming electron energy) one has the scaling:
\begin{align}
\label{eq:bremvsPrim}
    \sigma_{ae} \propto \aem^2 \ge^2\frac{m_e^2}{\ma^2} \,, \\
    \label{eq:Prim}
     \sigma_{a\gamma} \propto \aem \ga^2  \,, 
\end{align}
so that the bremsstrahlung cross-section is suppressed with respect to Primakoff production 
by the ratio $m_e^2/\ma^2$, by an extra factor of $\aem$, and by   
an additional suppression factor  from three-body vs. two-body phase space that is 
left implicit in Eqs.~\eqref{eq:bremvsPrim}, \eqref{eq:Prim}.
It is also worth noting that, for equal values of the masses
and up to order one factors from the different nature of the 
outgoing particles,  under the exchange: 
\begin{align}
    e \varepsilon \leftrightarrow \ge m_e \,,
\end{align}
the cross section for ALP production via bremsstrahlung  is equivalent to that 
for  bremsstrahlung production of  dark photons.
This reflects a generic feature that will prove to be quite useful in our work: the 
electron $t$-channel  ALP production closely  mimics the results for the dark photon. 

The  number of  ALP produced in a given electron beam dump experiments is  given by:
\begin{align}
\label{eq:NoeTracklength}
    \mathcal{N} = \frac{\mathcal{N}_AX_0\, \rho}{A} \int_{E_c}^{E_0} dE \ \frac{\partial T_{e/\gamma}}{\partial E}  \int_{E_{c}}^{E} dE_a \frac{d \sigma}{dE_a} ~\equiv~ \frac{\mathcal{N}_AX_0\, \rho}{A} ~\sigma_{\rm eff}\,, 
\end{align}
where we have accounted for the possibility of an experimental energy cut-off $E_c$ on the energy $E_a$ of the emitted ALP, 
and  $\mathcal{N}_A=6\times 10^{23}\,$mole$^{-1}$ is the Avogadro number, 
$A$ is the atomic number of the dump medium in g/mole, $X_0$ its 
radiation length in cm,  
 $\rho$ its mass density in $\rm{g} \, \rm{cm}^{-3}$,
$T_{e/\gamma}$ the electron or photon track length, and 
  the cross-section $ \sigma_{\rm eff}$  goes in 
  units of 
  $\rm{cm}^{2}$. In the last equality 
  we have defined an ``effective'' cross-section, which includes the integration over 
  ${\partial T_{e/\gamma}}/{\partial E}$. These differential track lengths represent the
 total length ( in unit of radiation length) 
of material traversed by all $e^\pm$ or $\gamma$ of a given energy $E$ 
 present in the shower.
 While the track lengths can be obtained directly via a Monte Carlo (MC)
 simulation of the shower, a relatively reliable expressions for electron 
 beam dump has been derived long ago in~\cite{Tsai:1966js}:
\begin{align}
\label{eq:tracklenghtTsai}
 \frac{\partial T_{e/\gamma}}{\partial E} = \int_{0}^{t_{tar}} \! dt \ I_{e/\gamma} (t,E), 
 \ \ \rm{ where } \quad \begin{cases} \  
    I_{e} (t,E) = \displaystyle \frac{1}{E}\frac{\left[\ln(E_0/E))\right]^{4t/3-1}}{\Gamma \left(4t/3\right) } \\[1em]
 \ I_{\gamma} (t,E) = \displaystyle \frac{1}{E}\frac{(1-E/E_0)^{4t/3}-e^{-7t/9}}{\frac{7}{9}+\frac{4}{3} \ln(1-E/E_0)}, \\
   \end{cases} 
\end{align}
where $t_{tar}$ is the target length in unit of radiation length. 
We have simulated both production processes in \amc using an effective $N-\gamma$ interaction with form factor $G_2$. This implies that we did not use the Weizsacker-Williams approximation for the cross section~\cite{Tsai:1986tx}, but we have instead estimated 
directly the $2 \to 3$ bremsstrahlung and $2 \to 2$  Primakoff processes. Furthermore, in order to regulate the numerical divergence which arises for large electron energies 
when the exchanged photon is very soft, we have modified the form factor $G_2(t)$:
we know that, due to the screening effects occurring when $a^2 t \ll 1$, this part of the phase space is sub-dominant in the final production rate. We have therefore imposed a regularising cut by setting the form factor to $0$ in the ``screened'' region:
\begin{align}
G_2^r (t) = \begin{cases} G_2 (t) &\text{   for } a^2 t > 1/3 \\ 0 &\text{   for } a^2 t < 1/3\ .
 \end{cases}
\end{align}
We have explicitly checked that the value of the  cross section is not modified by varying the cut between $a^2 t < 1$ and $a^2 t  < 0.05$, and that it agrees with the analytical expression developed above. Furthermore, we have verified that the differential distribution in angles and energy are also not affected by this regularisation. 

\begin{figure}[ht]
\begin{center}
        {\includegraphics[height=0.5\linewidth]{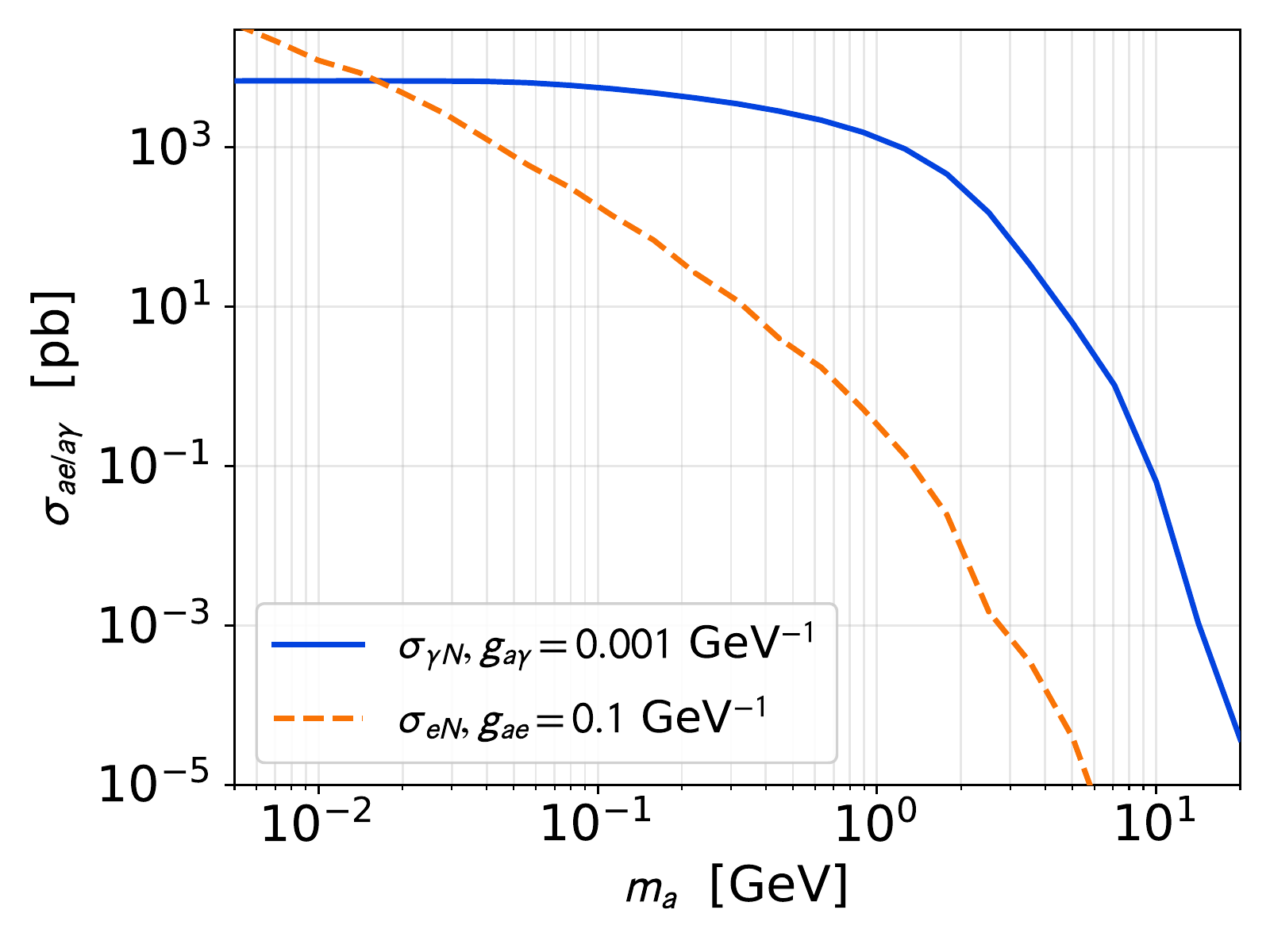}}
        \caption{Effective production cross-section for the Primakoff (solid blue line) 
        and bremsstrahlung (dashed red line) processes in NA64. The initial electron/photon energy is weighted by the corresponding track length distribution
            ${\partial T_{e/\gamma}}/{\partial E} $ 
        in the electromagnetic shower. }
\label{fig:csprodNA64}
\end{center}
\end{figure}

We conclude this section by illustrating in Fig.~\ref{fig:csprodNA64} the hierarchy between both production modes in the case of the NA64 beam dump (that will be  
thoroughly studied in the next section). As expected, the photon-driven production strongly dominates the total rate, even though photons are secondary particles in the 
$e^-$-initiated electromagnetic shower. This is due in part to the strong enhancement of the Primakoff production rate compared to bremsstrahlung, and in part from the fact that bremsstrahlung photons carry a large fraction of the energy of the scattered primary 
electrons and, being the NA64 target around $40$ radiation lengths
all primary electrons will  eventually undergo such a process.

\section{Experimental searches for invisible ALPs}
\label{sec:accelerators}
 
A trademark of the scenario  which is the subject of this work is that,
in contrast with standard long-lived ALP searches, here the invisible channel 
is assumed to dominate the ALP decay in all regions of parameter space. 
The signatures to search for such an ALP can be divided between ``pure'' missing energy searches, such as the ones performed by NA64, where one directly triggers on 
very low energy events created from high-energetic primaries, and ``associated'' 
missing energy searches, where the experiments use a single photon produced 
in association with the ALP to search for bumps on its recoil mass.

\subsection{Limits from $\mathbf{e^+ e^-}$ experiments}
\label{sec:analysis}

A first class of experiments rely on $e^+ e^-$ colliders to produce the ALP via the
process $e^+ e^- \to a \gamma$, and trigger on single photon events with a large missing energy from the escaping ALP. We will consider the re-interpretation of searches carried out in the BaBar~\cite{Lees:2017lec} and DELPHI~\cite{Abdallah:2008aa} experiments, as well as projections for the Belle-II sensitivity~\cite{Kou:2018nap}.  Due the ALP decaying mainly invisibly, the LEP constraints~\cite{Jaeckel:2015jla} on the visible 
mode $e^+e^-\rightarrow \gamma\gamma$ from are not relevant here.

As was discussed in the previous sections, the $\ge$ and $\ga$ couplings provide 
two  diagrams for  ALP production that, due to the suppression of  
interference effects, to a very good approximation correspond to 
two independent production processes. 
We can then define the total production rate times the experimental efficiency $\effCStot$ 
as the sum of the corresponding partial cross sections:
\begin{align}
   \effCStot = \ge^2 (\varepsilon \sigma )_{e} + \ga^2 (\varepsilon \sigma)_{\gamma} \ ,
\end{align}
where we have put in evidence the $\ga$ and $\ge$ dependence of the two contributions.
Since all  the experiments that we will consider follow a ``cut-and-count'' search strategy, 
we can estimate the limits $ \ge^{\rm{lim}}$ and $\ga^{\rm{lim}}$ from the two 
production process independently, and then combine both rates to obtain 
coupled limits as 
\begin{align}
\label{eq:combinedgage}
(\ge,\ga)^{\rm{lim}} \leqslant \left( \left(\frac{\ge^1}{\ge^{\rm{lim}}}\right)^2 + \left(\frac{\ga^1}{\ga^{\rm{lim}}}\right)^2\right)^{-1/2} \times (\ge^1,\ga^1) \, ,
\end{align}
where the pair of  couplings $(\ge^1,\ga^1)$ is used 
as an input to fix a priori a specific ratio between the two independent 
couplings. We will use the above approach for the DELPHI and PADME limits, for which we have fully simulated the expected signal. For BaBar (and for the Belle-II projection), 
we will instead rely on comparing directly $\effCStot$ 
with the limits on dark photon searches reported (projected) by these two experiments.

\paragraph{LEP-DELPHI.}

The DELPHI experiment at LEP has performed a single-photon search~\cite{Abdallah:2008aa} for 
the emission of an invisible graviton 
from low-scale extra-dimension. 
 Starting with Ref.~\cite{Fox:2011fx}, the results of this search have been  
 widely recasted to constrain  more generic types of invisible particles. 
Using $\calchep$\cite{Belyaev:2012qa} we have simulated $10^4$ $e^+e^-\rightarrow \gamma + a$ events, and we have followed~\cite{Abdallah:2008aa}  and \cite{Fox:2011fx} in 
imposing the following cuts 
\begin{align}
   45^{\circ}<\theta_{\gamma}<135^{\circ}\ \  \textrm{ and }\ \  0.9 <E_{\gamma}/E_{beam}<1.05 \ ,
\end{align}
where $\theta_{\gamma}$ ($E_\gamma$) is the photon angle (energy) in the center-of-mass frame. Note that there is no correlation between $E_{\gamma}$ and $\theta_{\gamma}$, as is clear analytically since $E_{\gamma}=(s-\ma^2)/(4E_{beam})$. The DELPHI experiment quotes an energy resolution~\cite{Abdallah:2008aa}: 
\begin{align}
\frac{\sigma_E}{E_{\gamma}}=0.0435 \oplus \frac{0.32}{\sqrt{E_{\gamma}}} \ ,
\end{align}
where $\oplus$ denote the sum of two independent Gaussian distributions. This leads to a significant smearing of an otherwise narrow  distribution centered around $E_{\gamma}=(s-\ma^2)/(4E_{beam})$, and is particularly important to model properly the threshold in $\ma$ induced by the energy cuts $0.9 <E_{\gamma}/E_{beam}<1.05$. Altogether, these cuts lead to a selection efficiency of around $40 \%$ for, e.g. $\ma=10$ MeV and $\ga=0.1$ GeV$^{-1}$.

\begin{table}[h!]
\begin{center}
\small
\begin{tabular}{|c|c c c c c c c c c c |}
        \hline   \rule{0pt}{1.25em}  $\langle \sqrt{s} \rangle$ (GeV)& $182.7$ & $188.6$  & $191.6$ & $195.5$ & $199.5$ & $201.6$& $203.7$ & $205.2$& $206.7$ & $208.2$\\[0.5em]
         \hline
        \rule{0pt}{1.25em}    $\mathcal{L}$ [pb$^{-1}$]& $50.2$ & $154.7$& $25.9$& $76.4$& $83.4$& $40.6$& $8.4$& $76.2$& $121.6$& $8.3$ \\[0.3em] \hline 
                \end{tabular}
        \caption{Luminosity acquired in the DELPHI HPC detector as function of the average CoM energy delivered by LEP~\cite{Abdallah:2003np}.}
\label{tab:lumiDELPHI}
\end{center}
\end{table}

\begin{figure}[h]
\begin{center}
        \includegraphics[height=0.5\linewidth]{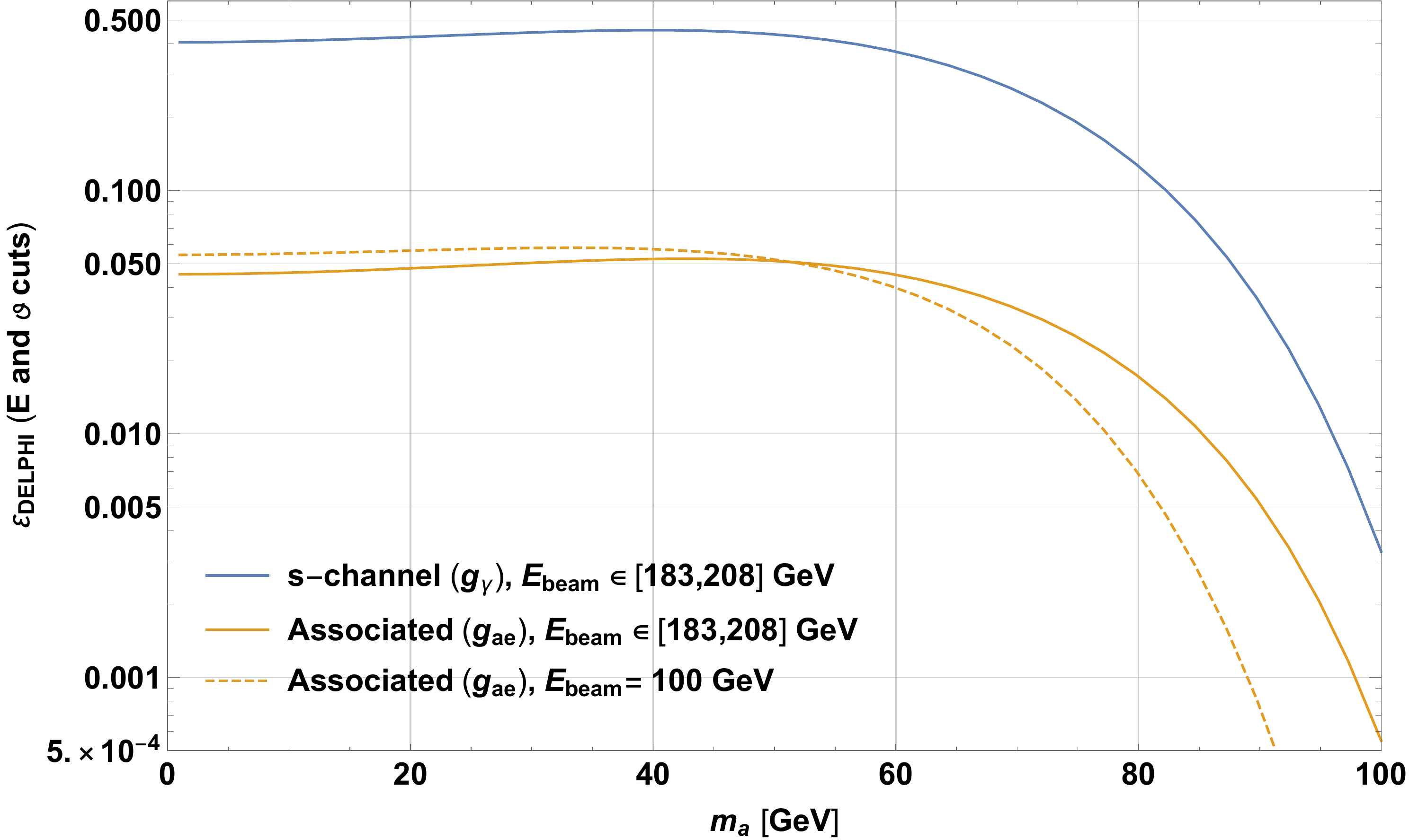}
        \caption{Estimated total  efficiencies for the energy cut $0.9 <E_{\gamma}/E_{beam}<1.05$ and angular cut $45^{\circ}<\theta_{\gamma}<135^{\circ}$ at DELPHI, for the $s$-channel ALP production process controlled by $\ga$ 
        (orange continuous line)
        and for the associated production controlled by $\ge$ (blue line).
        The dashed orange line correspond to the efficiency for the associated production in the case the CoM energy is fixed at the average value 
         $\langle \sqrt{s}\rangle  = 200\,$GeV.}
\label{fig:effDELPHI}
\end{center}
\end{figure}

The full DELPHI data sample was collected  at different CoM energies,  
ranging from $180.8\,$GeV to $209.2\,$GeV~\cite{Abdallah:2003np}. The corresponding accumulated luminosity per energy bin is given in Table~\ref{tab:lumiDELPHI}. Although the 
change in CoM energy  was not relevant for Ref.~\cite{Fox:2011fx}, which considered 
off-shell new physics operators whose effects were spread in the various energy bins, 
in our case it is important to model properly the various mass thresholds. 
We have therefore estimated the efficiency for the energy cuts as a weighted sum over the integrated luminosity per energy bin. Finally, 
following~\cite{Fox:2011fx}  we have also included an overall trigger efficiency
 $ \varepsilon_{t} = 84\%$. This yields the full efficiency:
\begin{align}
    \varepsilon_{\rm DELPHI} = \varepsilon_{\theta} \times \varepsilon_{t} \times \sum_i \frac{\mathcal{L}_i}{\mathcal{L}_{\rm tot} } \varepsilon_{s_i} \ ,
\end{align}
where $\varepsilon_{s_i}$ is the efficiency of the energy cuts for a LEP CoM energy $\sqrt{s_i}$ and luminosity $\mathcal{L}_i$ as given in Table~\ref{tab:lumiDELPHI}, $\varepsilon_{\theta}$ is the efficiency of the angular cut, which depends on the ALP production process, and $\mathcal{L}_{\rm tot} = 645.7 \textrm{pb}^{-1}$ is the total integrated luminosity.
The efficiencies of the angular and energy cuts for the two main ALP production mechanisms
are plotted in Fig.~\ref{fig:effDELPHI}.
The dashed orange line represents the efficiency for the  $t$-channel associated 
production assuming that the entire luminosity is collected  at the 
average CoM energy $\langle \sqrt{s}\rangle  = 200\,$GeV, 
while the orange  solid  line corresponds to  the more accurate result obtained 
by accounting for the CoM energy of each partial integrated luminosity.
We see that while there is good agreement for small ALP masses, the results 
differ significantly for an ALPs mass exceeding $m_a\sim 70-80\,$GeV. 
The solid blue line in the figure depicts the complete result for  
 the photon $s$-channel process. It can be seen that the overall efficiency 
 for this channel is much higher than that for the associated production. 
This is due to the angular cuts, and reflects the different shape of the angular 
differential distribution that, for the associated production, is peaked at small angles.
For the statistical analysis of ALP production we need to compare the simulated 
signal events with the collected data. One additional complication is that, 
especially in the last bin, the agreement  between the predicted number 
of background events from the MC simulation in~\cite{Abdallah:2008aa} and the actual data
is rather poor.
However, the simulation of $e^+ e^- \to \gamma \nu \bar{\nu}$ events performed a posteriori in~\cite{Fox:2011fx} predicted around $49$ such background events for $E_\gamma/E_{\rm beam}$ 
in the interval $[0.9,1.05]$, in better agreement with the $61$  events observed. 
Altogether, a first conservative approach is to assume that every observed event originates from an ALP, setting the limit on the number of observed events to $N^{90}_{lim} \sim 72$ at 90\% confidence level (C.L.). A more aggressive approach is instead to account for the $e^+ e^- \to \gamma \nu \bar{\nu}$ simulation from~\cite{Fox:2011fx}, use an overall $10 \%$ uncertainty on this background
(to be compared to the $5 \%$ systematic uncertainty quoted in~\cite{Abdallah:2003np}), and to assume that the remaining events are produced by ALPs. This
leads instead to $N^{90}_{lim} \sim 25.5$ at $\clsnf$, and to a $\sim 40\%$ stronger 
limit with respect to the conservative approach.
In the next sections, when providing the final limits, we will adopt the 
latter choice,
keeping in mind that this 
stems from somehow aggressive assumptions. 
Once $ \ge^{\rm{lim}}$ and $\ga^{\rm{lim}}$ have been obtained for both ALP production processes as function of the ALP mass $\ma$, they  can eventually be combined 
as in Eq.~\eqref{eq:combinedgage}.

\paragraph{BaBar.}
The BaBar experiment at the PEP-II B-factory has accumulated a large dataset  of $e^+ e^-$ collision events, corresponding to a total luminosity of 53 fb$^{-1}$. Single photon events with large missing energy were analysed via a bump-search~\cite{Lees:2017lec}, focusing on the case of a dark photon.\footnote{The BaBar collaboration had previously  carried out  also a search for production of invisible final states in single-photon decays of $\Upsilon(1S)$~\cite{delAmoSanchez:2010ac},
which however yields limits that are less stringent than those implied by their more recent search~\cite{Lees:2017lec}  
on which we focus here.}

\begin{figure}[h]
\begin{center}
        {\includegraphics[width=0.49\linewidth]{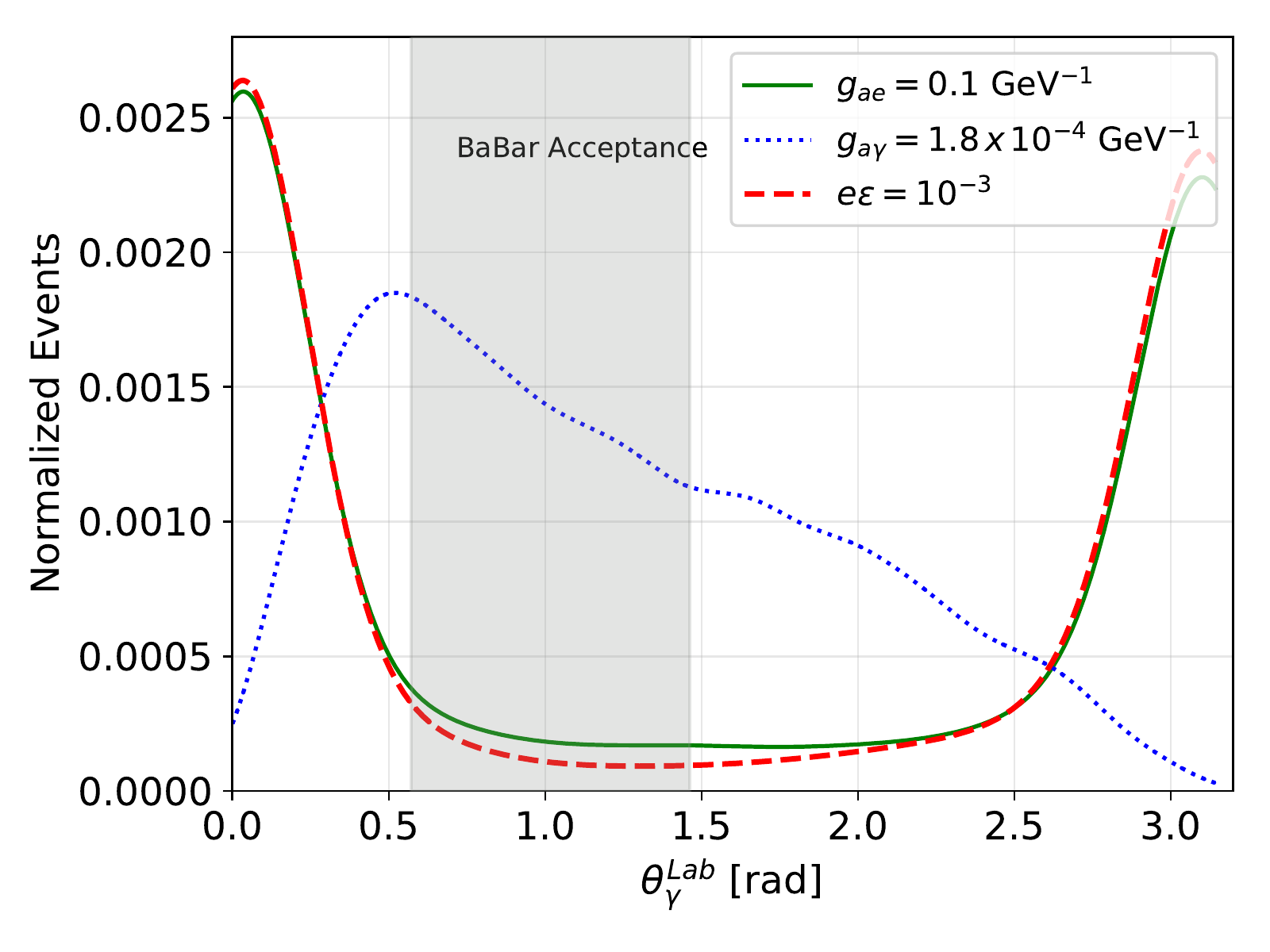}}\hspace{0.05cm}
        {\includegraphics[width=0.49\linewidth]{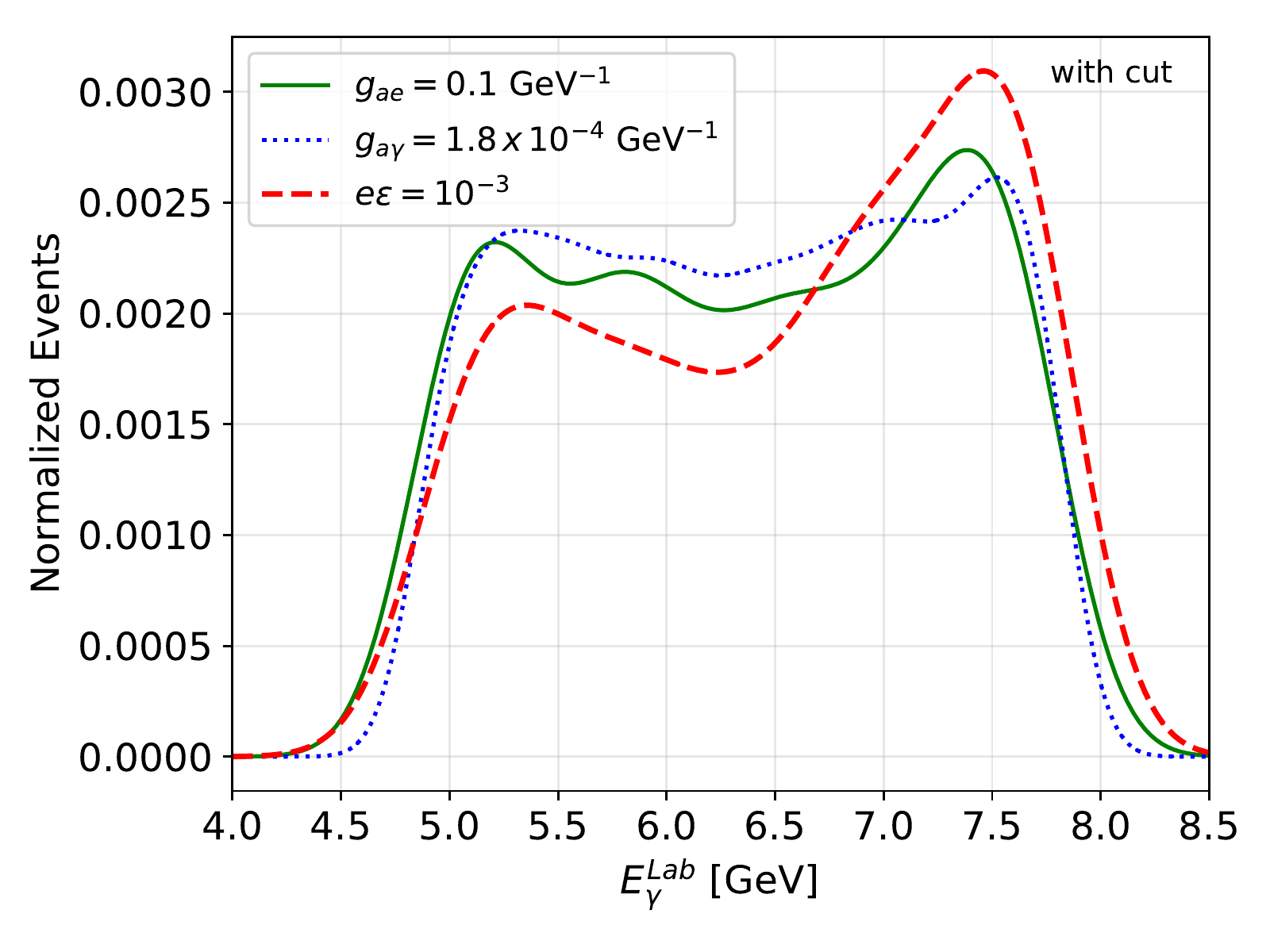}}
        \caption{Normalized photon emission angle distribution (left panel) and energy 
        distribution with BaBar angular cuts (right panel)  
        in the laboratory frame for $e^+e^-\rightarrow \gamma X$  with 
           $X=\text{dark photon, ALP}$ and 
        $M_X= 20\,$MeV. The red dashed line is for 
        a dark photon with a coupling $e \epsilon=10^{-3}$, the green line for an ALP 
        with $\ge=10^{-3}$ GeV$^{-1}$, and the blue dotted line for an ALP with 
        $\ga=1.8\times10^{-4}$ GeV$^{-1}$. The grey region represents the BaBar angular coverage.}
\label{distrbabar}
\end{center}
\end{figure}

As such, the limits reported by the collaboration need to be modified to account for the different production channels and kinematics of the ALP case.
We have use the \calchep simulation setup described above, tuned to the BaBar characteristics: an asymmetric interaction between $9$ GeV electrons and $3.1$ GeV positrons to produce a CoM collision energy of $10.58$ GeV. The selection cuts on the CoM photon angle 
\begin{align}
-0.4<&\text{cos}\,\theta_{\gamma}<0.6  & \textrm{ for } \ma < 5.5 \textrm{ GeV}\,\, \\
-0.6<&\text{cos}\,\theta_{\gamma}<0.6  & \textrm{ for } \ma > 5.5 \textrm{ GeV} \,, 
\end{align}
were applied directly to the MC truth events  using Root~\cite{Brun:1997pa}. Note that BaBar selection cuts further require $E_{\gamma} > 3 ~(1.5)$ GeV for the low (high) mass search, but this cuts are redundant because of the angular ones up to  
a threshold mass of about $\ma \sim 8 \textrm{ GeV}$. We plot in Fig.~\ref{distrbabar} the energy and the emission angle distributions for  a dark photon and 
for the two  ALP production processes. As expected, the two cases have a priori very distinct differential distributions. Note that since the photon energy in the CoM frame is fixed as $E^{CoM}_{\gamma}=(s-\ma^2)/(2\sqrt{s})$, the correlation between the energy and the angle 
in the laboratory frame
is of purely kinematic origin, and does not depend on the nature of the emitted particle. 
On the other hand, as can be seen from Fig.~\ref{distrbabar}, the efficiencies of the geometric cuts  differs strongly for the two cases and need to be fully accounted for. Note  that the full BaBar analysis uses a Boosted-Decision-Tree (BDT) approach, whose efficiency as function of the signal characteristics ($E_\gamma$ and $\theta_\gamma$) is not reported. However, as is clear from Fig.~\ref{distrbabar}, the actual distribution of signal events within the angular cuts differ (up to an overall normalisation factor) by at most $25\%$. 
Since the BDT only considers the signal on an event-per-event basis, 
 a BDT variation in efficiency between the two cases cannot be significantly 
 larger than this number, which  is then included as an additional uncertainty in our final result.
In order to extend the limits from~\cite{Lees:2017lec} to the case of an ALP with both $\ga$ and $\ge$ couplings, we have calculated the cross-section times efficiencies for $e^+ e^- \to \gamma a$ in the acceptance region of BaBar as function of the ALP mass. We then estimate  the ratio of this quantity  
with the corresponding one for the production of a dark photon $V$ of 
the same mass $m_V=\ma$ and for a kinetic mixing $\varepsilon = 10^{-3}$:
\begin{align}
    R (\ga,\ge,\ma) ~\equiv~ \frac{ (\epsilon \sigma)_e + (\epsilon \sigma)_a}{ (\epsilon \sigma)_{e^+ e^- \to \gamma V}}\,. 
\end{align}
The limit on the ALP couplings is then given by:
\begin{align}
    (\ge^{\rm{lim}},\ga^{\rm{lim}}) = \left(\frac{\varepsilon^{\rm{lim}}_{\rm{BaBar}}}{0.001} \right)\frac{1}{\sqrt{R (\ma)}}\,, 
\end{align}
where $\varepsilon^{\rm{lim}}_{\rm{BaBar}}$ is the limit reported by the collaboration.  
The final result only depends on the ratio between the ALP couplings $\ge/\ga$ so that the ratio $R$ can be  evaluated 
by keeping  the sum of the square of 
the couplings fixed to some specific value 
$\ge^2 + \ga^2 = \bar g^2$.

\paragraph{Prospects for Belle-II.}

The experiment Belle-II~\cite{Kou:2018nap} is currently accumulating data and 
in the coming years is expected to produce  significantly stronger limits than BaBar. 
Projections for mono-photon searches based on the same guiding principles as in BaBar 
and based on a fast data acquisition of $20 \,\rm{fb}^{-1}$ and a possible final experimental dataset of $50 \,\textrm{ab}^{-1}$ have been given in Refs.~\cite{Dolan:2017osp,Kou:2018nap}.
We have recasted the projections by adopting the same approach used for BaBar. We 
estimated the expected number of events for both the dark photon and the ALP, including the following experimental cuts (in the CoM frame):
\begin{align}
    -0.85 < \cos \theta_\gamma < 0.91 \ ,
\end{align}
where we used the boost factor from laboratory to CoM frame at Belle-II of $0.28$ (compared to $0.56$ for BaBar) and the angular coverage of the triggers from Ref.~\cite{Kou:2018nap}. Similarly to the BaBar case we do not include directly the energy cuts since their effect is mostly included in the limits shown in~\cite{Kou:2018nap}. Interestingly, due to the larger angular coverage, the kinematics differences between ALP-related $s$-channel process and the dark photon one are somewhat larger than for BaBar (with dark photon distributions strongly localised around the upper limit for the mono-photon energy). This implies that 
an accurate estimate of Belle-II limits on $\ga$ would in fact require a dedicated simulation. 
Note however, that our projection still agrees very well with the 
estimate of the Belle-II reach performed in Ref.~\cite{Dolan:2017osp}. The same 
issue does not arise in estimating the Belle-II reach for $\ge$, as 
in this case the signal broadly shares the same characteristics than 
the dark photon signal.

\subsection{Limits from electron beam dump experiments}

Let us now focus on the limits arising from past and current electron beam dump experiments. Standard searches typically assume the ALP to decay visibly, and we thus expect limits from past electron beam dumps such as from the SLAC E141~\cite{Riordan:1987aw} or  E137~\cite{Bjorken:1988as} experiments to be  weakened strongly. In the following we will describe quantitatively  this weakening by studying the E137 limits as function of the ALP visible branching ratio. Conversely, searches  at NA64~\cite{Banerjee:2020fue} do not always 
rely on signals from visible decays, and will accordingly lead to stronger constraints.

\paragraph{E137.}

The SLAC E137~\cite{Bjorken:1988as} experiment searched for light ALPs using a $20$ GeV  electron beam dumped into aluminium (later iron) plates interlaced with cooling water. The shielding was provided from a neighbouring hill, corresponding to $D=179$ m of shielding, and was followed by $L=204$ m of open air decay region. The electromagnetic calorimeter was $2\text{m} \times 3\text{m}$ ($3\text{m} \times 3\text{m}$ ) for Run 1 (Run 2) and could detect charged particles and/or photons. The experiment accumulated a total dataset of $1.86 \times 10^{20}$ electrons on target (EoT),  one third during the first run 
(with $\sim 10\,$C of total electron charge) and the rest during Run 2 ($\sim 20\,$C).
No candidate events with an energy above $\sim 2 $ GeV and satisfying the various angular and timing cuts were found. Although one event was observed with energy higher than $1$ GeV, there is the possibility that it originated from a cosmic ray muon~\cite{Batell:2014mga}. 

In presence of both $\ge$ and $\ga$, the ALP will be predominantly produced by Primakoff effect, due to the suppression of the bremsstrahlung process shown in Eq.~\eqref{eq:bremvsPrim}.\footnote{Note that the presence of secondary positrons in the electromagnetic showers implies that resonant production of ALP can also occur, although 
with a reduced geometric acceptance~\cite{Marsicano:2018krp}.} Hence  in the following we will focus exclusively on this production mechanism. Since  the E137 search 
assumed visible decays of a long-lived ALP, it is clear that the reported limits get  significantly weakened in a scenario in which invisible decays dominate. In order to estimate this effect, we start by considering the recasting of Ref.~\cite{Dolan:2017osp} which 
inferred  an upper limit on the couplings a function of the ALP mass
for short-lived ALPs, on the basis of  the single, 
lower-limit given by the experimental collaboration~\cite{Bjorken:1988as}.

The E137 search for a long-lived ALP excludes a region of 
ALP-photon couplings between a lower limit $g_d^{\rm E137}$ corresponding to a very long-lived ALP  and an upper limit $g_u^{\rm E137}$ corresponding to a short-lived ALP. Both limits are function of the mass $\ma$, and ALPs with couplings $\ga \in [g_d^{\rm E137} , g_u^{\rm E137}]$ are excluded since they would decay in front of the apparatus with experimentally excluded rates. 
$g_d^{\rm E137}$ and $g_u^{\rm E137}$ 
have been extracted from Ref.~\cite{Dolan:2017osp} (see in particular Fig. 2). If additionally the ALP has a significant invisible branching ratio, the upper limit will be significantly modified, since the invisible decay channels contribute  to reduce the ALP 
lifetime. On the other hand, the lower limit, corresponding to the case of very long-lived ALP, does not get modified. In order to estimate the new upper limit $g_u^{\rm new}$ we solve 
\begin{align}
\label{eq:recastguE137}
    (g_u^{\rm new})^2 \, \mathcal{P}^{\rm new} = (g_u^{\rm E137})^2 \ \, \mathcal{P}_u^{\rm old} \ ,
\end{align}
where $ \mathcal{P}^{\rm new}$  and $ \mathcal{P}^{\rm old}$ 
denote respectively the ALP decay probability with and without  invisible decay channels, and are given by 
\begin{align}
\label{eq:Pnew}
    \mathcal{P}^{\rm new} &\equiv~ \exp \left( {\displaystyle - \frac{(L+D)\Gachi (g_u^{\rm new})}{c \hbar \gamma_a}} \right) \ \frac{L \Gavis (g_u^{\rm new})}{c \hbar \gamma_a} \\[0.8em]
    \label{eq:Pold}
    \mathcal{P}^{\rm old} &\equiv~ \exp \left(\displaystyle   ~- \frac{D\Gavis (g_u^{\rm E137})}{c \hbar \gamma_a}\right) \ . \
\end{align}
In these relations we have assumed that the detection and geometric efficiencies are solely a  
function of the ALP mass, we have approximated $\Gamma_{\rm tot} \approx \Gachi$ for decays in the shield of length $D$, 
and we have included the ALP boost factor $\gamma_a = E_a/ \ma$.

The above formulas further uses the fact that: (1) if the ALP decays mainly invisibly, the visible decay probability is small enough to approximate the exponential term by $\sim L \Gavis (g_u^{\rm new})$; (2) in the original case of a visibly decaying ALP, the decay probability for the upper limit only  requires that the ALP has not decayed in the shielding. Indeed, since $L \gtrsim D$, the 
surviving ALPs will then decay in the decay volume with near $100\%$ probability.

Altogether, it is clear that the only missing information in these expressions 
is the ALP boost factor, which depends on the ALP energy distribution after 
geometrical cuts are applied. An average value 
for $\gamma_a$ can in fact be extracted using the above relations and 
the upper and lower limits $g_u^{\rm E137}$ and $g_d^{\rm E137}$. We obtain:
\begin{align}
    \hbar c \langle \gamma_a \rangle = D \, \Gagam(g_u^{\rm E137})  \left[  \mathcal{W} \left( \frac{(g_u^{\rm E137})^4 D }{ (g_d^{\rm E137})^4 L}\right)\right]^{-1} \ ,
\end{align}
where we have used the Lambert $\mathcal{W}$ function, and $\Gagam(g_u^{\rm E137})$ is the ALP decay width to di-photon evaluated for $\ga = g_u^{\rm E137}$. For small ALP masses, we find $m_a \langle \gamma_a \rangle \sim 15$ GeV, in accordance with the naive expectation based on the
fact that the  E137 electron beam energy is $20$ GeV.\footnote{Note that in all beam dump experiments, the upper limits are typically dominated by the fraction of events with the highest boost  since, although for  lower boosts one expects larger production 
rates, the number of detectable events gets exponential suppressed because of 
early decays.} Replacing now 
in Eqs.~\eqref{eq:Pnew} and \eqref{eq:Pold}
 $\gamma_a$ with   $\langle \gamma_a \rangle$  
 and using  Eq.~\eqref{eq:recastguE137},
 we can readily find the new upper limit $g_u^{\rm new}$.

Interestingly, the lower limit is not modified. This is because  for 
both cases of dominant visible or invisible decays, it corresponds to the  
regime where the ALP is long-lived.
Consequently, the number of expected visible decays simply scales with the 
visible width, as in the standard case. The above relations could also be 
applied to other beam dump limits on visible ALP decay 
(see, e.g.~\cite{Dobrich:2019dxc} for a recent update).

An important comment is that while the ALP may decay invisibly, it is clearly possible that its decay products still leave a signal in the detector. A typical example could be light dark matter scattering, such as in the case of the dark photon (note however, that the higher dimensional nature of the ALP portal tends to suppress this contribution w.r.t the dark photon case). Another possibility is a three-body decay of some heavier dark sector state into lighter ones, as is typically found in inelastic dark matter scenarios. Since in this work we do not specify the structure of the dark sector, we will not include any of such limits.  Finally, it has been pointed out recently that the inverse Primakoff process can also play a role to let light dark matter leave a signature in a detector~\cite{Brdar:2020dpr}.

\paragraph{NA64.}

 NA64  is a  fixed-target experiment 
which uses a~$100\,$GeV electron beam from the CERN SPS
secondary beam line H4, and combines active beam dump and 
missing energy techniques to search for rare events.
The experiment has accumulated $2.84 \times 10^{11}$ electrons-on-target (EoT) during the 2016-2018 run, and is expected to reach a total of  $5 \times 10^{12}$ EoT.  ALP production proceeds as in E137, with the Primakoff mechanism from secondary photon occurring  directly within the electromagnetic calorimeter (ECAL).
The ECAL is an active target assembled from lead and scandium plates 
that we model as $40$ radiation lengths dump of lead. 
The NA64 search for ALPs~\cite{Banerjee:2020fue} 
exploits two signatures:  $a\to \gamma\gamma$ with an energy deposition $E_{\rm{ECAL}} < 85\,$GeV, 
or large missing energy  with $E_{\rm{ECAL}}< 50\,$GeV. 
For our study the second signature is the relevant one, which implies that  
the ALP escaping ECAL needs to carry away at least $50\,$GeV 
of missing energy. 
No events were observed by the collaboration~\cite{Banerjee:2020fue}, in agreement with 
an expected background of around $0.19 \pm 0.07$. 

We have described in Sec.~\ref{sec:totxsecLO} a simple approximation to the ALP production rate that applies to the NA64 setup. Note that the requirement that the ALP carries 
at least half of the initial electron energy implies that only events in which an 
ALP is produced in the first stage of the development 
of the electromagnetic shower in the ECAL will be relevant for this search.
Including an experimental efficiency $\epsilon_{\rm{NA64}} \simeq 50\%$, we find 
that the expected number of events at NA64 is:
\begin{align}
 \label{eq:NA64}
 N_{ALP}=N_{EoT}\, \epsilon_{\rm{NA64}} \times \mathcal{N}_A 
 \frac{\rho_{\rm Pb}}{A_{\rm Pb}} 
 X_{0, \rm Pb} \times \left[ \left(\frac{\ge}{1 \textrm{ GeV}^{-1}} \right)^2 \sigma^{\rm eff}_{ae} + \left(\frac{\ga}{1 \textrm{ GeV}^{-1}} \right)^2 \sigma^{\rm eff}_{a\gamma} \right] \ ,
\end{align}
where $N_{EoT}$ is the number of electrons on target, 
 $\rho_{\rm Pb} = 11.4$g/cm$^3$ is the lead mass density, 
$A_{\rm Pb}=207\,$g/mole the atomic mass,  
$X_{0, \rm Pb} = 0.56\,$cm the radiation length, and 
the effective cross-sections $\sigma^{\rm eff}_{ae}$ and $\sigma^{\rm eff}_{a\gamma}$,  
defined according to Eq.~\eqref{eq:NoeTracklength}, 
include the track-lengths integration as detailed in Eq.~\eqref{eq:tracklenghtTsai}. 
Note that the effective cross-sections account for the interaction between the electrons 
and the target nuclei, so that they include factors of the atomic number $Z_{\rm Pb}=82$. 
For the  reference cross-sections $\sigma^{\rm eff}_{a\gamma}$ and $\sigma^{\rm eff}_{ae}$ 
we fix the couplings at the values $\ga=1 \textrm{ GeV}^{-1}$ and $\ge=1 \textrm{ GeV}^{-1}$, 
and then we use the explicit dependence on the couplings as given 
in Eq.~\eqref{eq:NA64} to derive the limits. 
To estimate the cross sections we have  used the numerical setup based on \amc~described in Sec.~\ref{sec:totxsecLO}.
 We have included directly at the MC truth-level the energy cut, requiring the ALP to carry away 
at least $50\,$GeV. We set the $\clsnf$ limit at $2.3$ ALP events deriving
limits on $\ge$ and $\ga$ as well as prospects for the full NA64 dataset of  $5 \times 10^{12}$ EoT. 
Remarkably, for the case of an ALP that is only  
coupled to  photons ($\ge=0$) and for low values of $\ma$, 
our estimate of the limit agrees within $5\%$ with the complete result 
from the collaboration.\footnote{The low mass limit corresponds to the case when the visibly decaying ALP 
searched for by the collaboration is long-lived, and hence mimics our signature, see also ref.~\cite{Dusaev:2020gxi}.}

\subsection{Simulation of ALP production in the PADME environment}
\label{sec:simulation}

The PADME experiment~\cite{Raggi:2014zpa} at LNF 
uses a positron beam from the DA$\Phi$NE LINAC accelerator in fixed target configuration, using as active  target $100\,\mu$m of polycrystalline diamond. The beam energy 
can be varied in the range $200-550$ MeV. The experiment aims to collect a sample of $N_{poT} \sim 4 \cdot 10^{13}$ positrons on target (poT) and has recently completed RUN II collecting  $\sim 5 \cdot 10^{12}$ poT. The current luminosity is mostly limited by 
a low duty factor $\sim 10^{-5}$ corresponding to 49 bunches of $200\,$ns length per second, each containing $\sim 2.8\cdot 10^4$ positrons.
Increasing the number of positrons per bunch would be easily feasible, but it would 
lead to an excessive  pile-up in the  experiment. Future experimental prospects include large improvements to the DA$\Phi$NE beam  serving the experiment, (POSEYDON proposal~\cite{Valente:2017hjt}), increasing the duty cycle in the  $10^{-2}-10^{-1}$ range by  spreading  the bunches over $0.2-2\,$ms
at the same frequency, which would allow to increase  $N_{poT}$ to around 
$ (4 - 40) \cdot 10^{16}$ while reducing pile-up in the experiment.
The electromagnetic calorimeter (ECAL) is designed to measure the final photon $4$-momentum. It is placed at
$3.45\,$m from the active target and it has a diameter of 60\,cm and consequently the basic cuts applied on the final-state are~\cite{Piperno:2020pvj}:
\begin{itemize}
\item{angular coverage: ($15\lesssim\theta_{\gamma}\lesssim 80$)\,mrad, with $\theta_{\gamma}$ the photon angle with respect to the beam direction  in the laboratory frame;}
\item{energy range for a reconstructed cluster in ECAL:  30\,MeV $\lesssim E_{\gamma}\lesssim$ 500 MeV.}\footnote{Note that for the associated production of ALP+$\gamma$, the 
energy of the emitted photon is always bwlow $500\,$ MeV for a beam energy 
of $550\,$MeV.}
\end{itemize}

\begin{figure}[t]
\begin{center}
\begin{subfigure}[b]{0.49\textwidth}
\includegraphics[width=\textwidth]{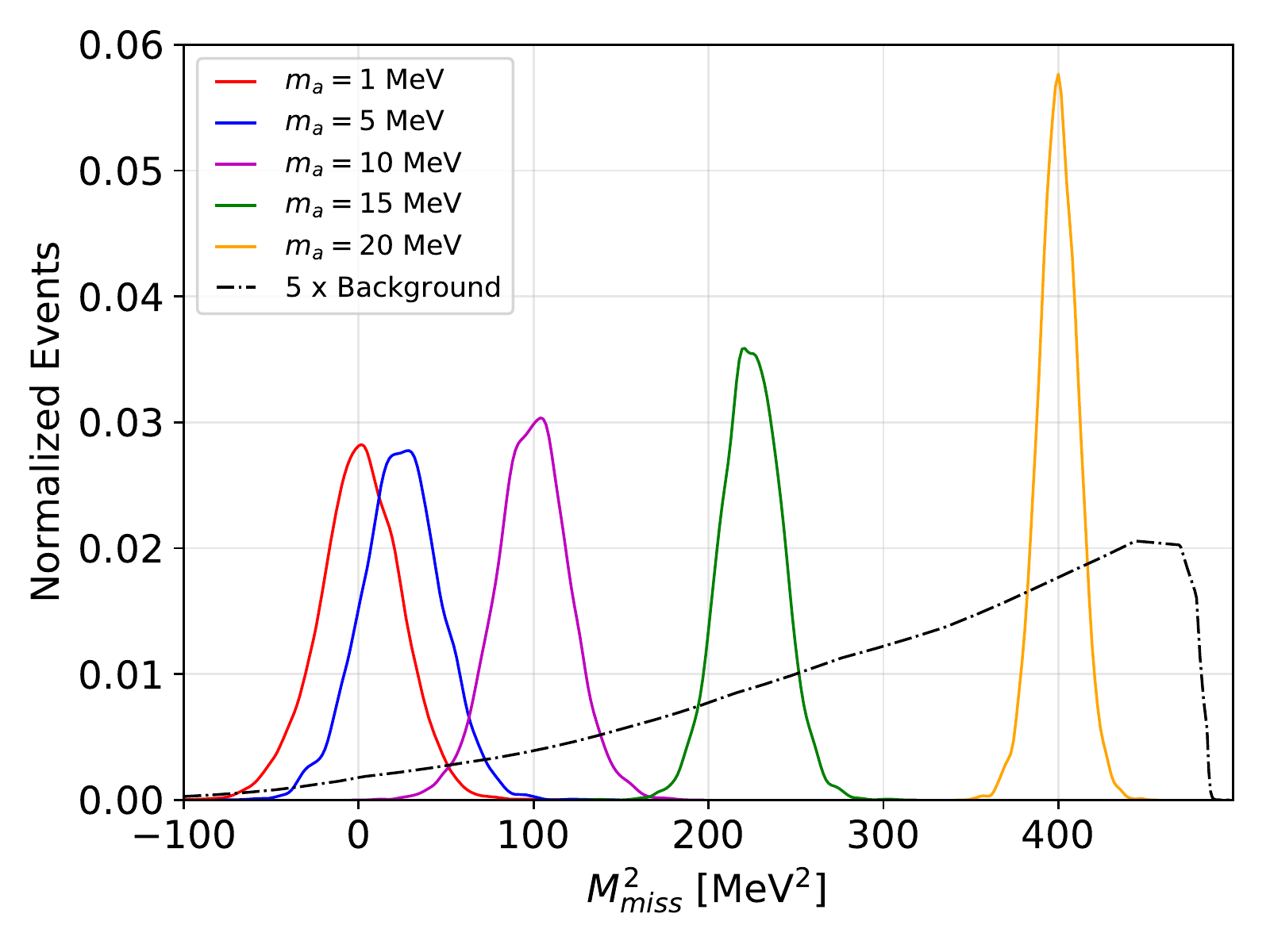}
\caption{ }
\label{MM2}
\end{subfigure}
\begin{subfigure}[b]{0.49\textwidth}
\includegraphics[width=\textwidth]{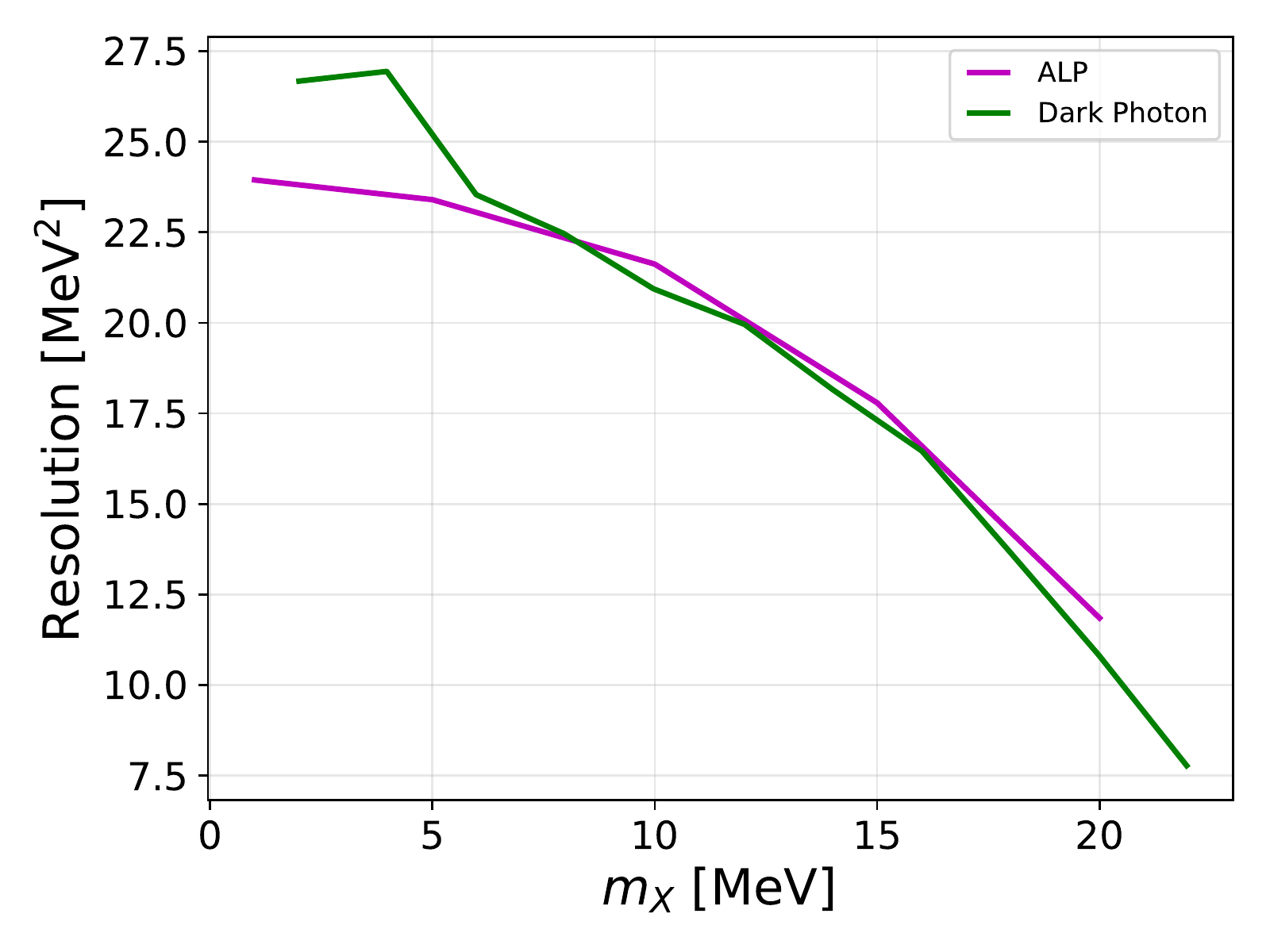}
\caption{ }
 \label{reso}
\end{subfigure} 
\end{center}
\caption{
(a) Normalized $M^2_{miss}$  distribution for a set of ALP masses 
with PADME cuts applied, including energy and spatial resolution. The normalized distribution of background events (dashed-dotted line) is extracted from \cite{Raggi:2014zpa}. 
(b) 
 $M^2_{miss}$  resolution as a function of the dark particle mass $m_X$, 
where $X$ is an ALP or a dark photon, for 
$E_{beam}=550\,$MeV.}
\label{fig:MM}
\end{figure}
The  experimental technique adopted by  PADME relies on the measurement of the missing mass in the final state. Indeed, knowing the initial conditions and measuring the single-photon four-momentum $p_{\gamma}$, it is possible to measure $\ma^2$ as the square 
of the missing mass:
$$M^2_{\rm miss}=(p_{e^+} + p_{e^-}- p_{\gamma})^2.$$
The signal would then correspond to a peak at $M^2_{\rm miss} = \ma^2$ over a smooth distribution from the background $e^+e^-\rightarrow \gamma\gamma(\gamma)$ and 
from bremsstrahlung.
Fig.~(\ref{MM2}) shows the   missing mass square distributions for different values of $\ma$ together  with an estimate of the  background based on GEANT4 MC simulation from Ref.~\cite{Raggi:2014zpa}. The expected number of background events is around 3800 through the selection of $10^{11}$ simulated events. For reproducing the missing mass distribution we have simulated $10^5$ events by \calchep and applied the PADME cut using Root. We have also accounted for the spatial resolution of $4\,$mm and the measured energy resolution of ECAL~\cite{Piperno:2020pvj}, 
\begin{align}
    \frac{\sigma_E}{E_{\gamma}}=\frac{2\%}{\sqrt{E_{\gamma}\text{[GeV]}}}\oplus\frac{0.03\%}{E_{\gamma}\text{[GeV]}}\oplus 1.1\%.
\end{align}
\begin{figure}[t]
\begin{center}
        {\includegraphics[height=0.35\linewidth]{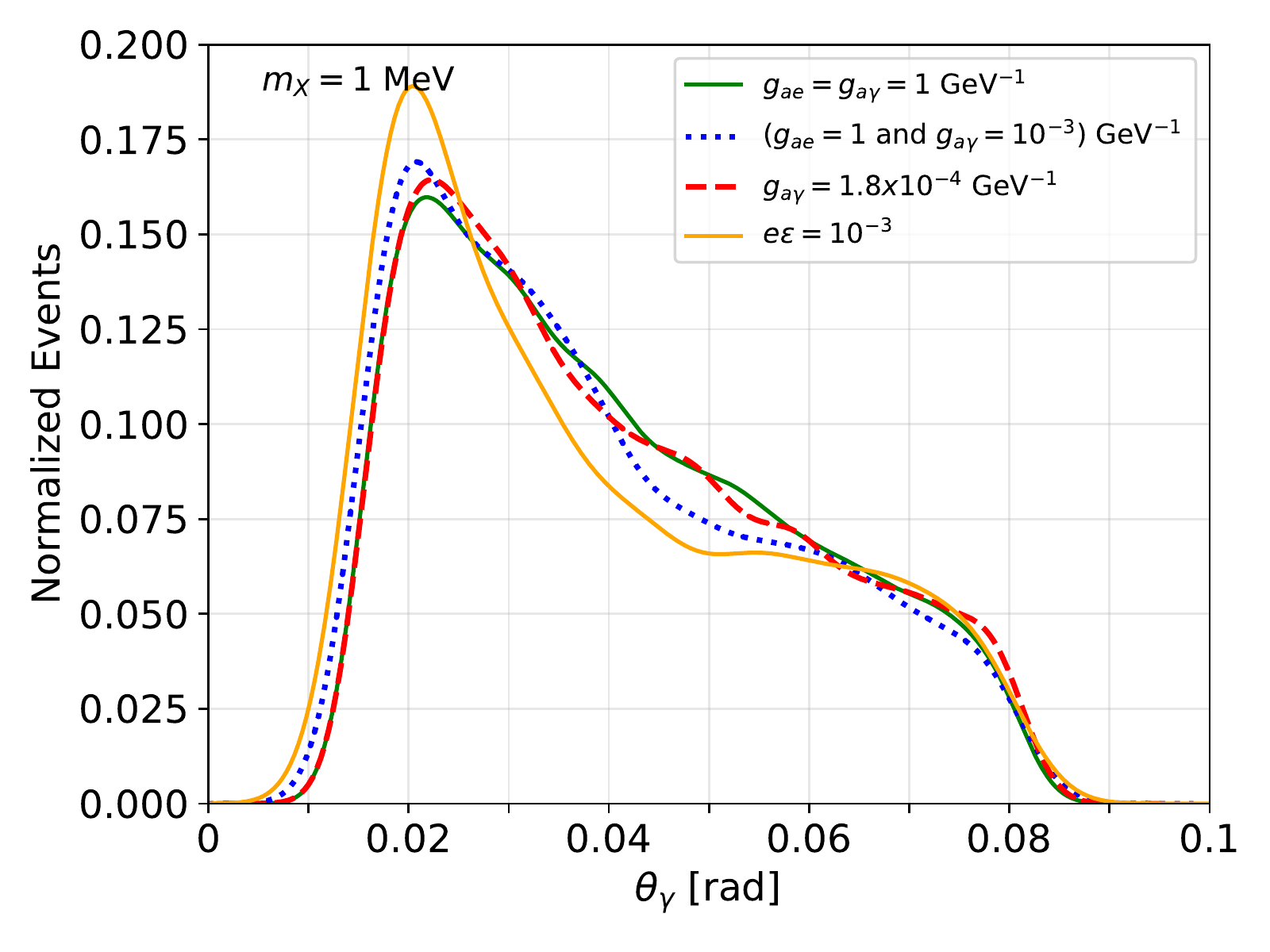}}\hspace{0.06cm}
        {\includegraphics[height=0.35\linewidth]{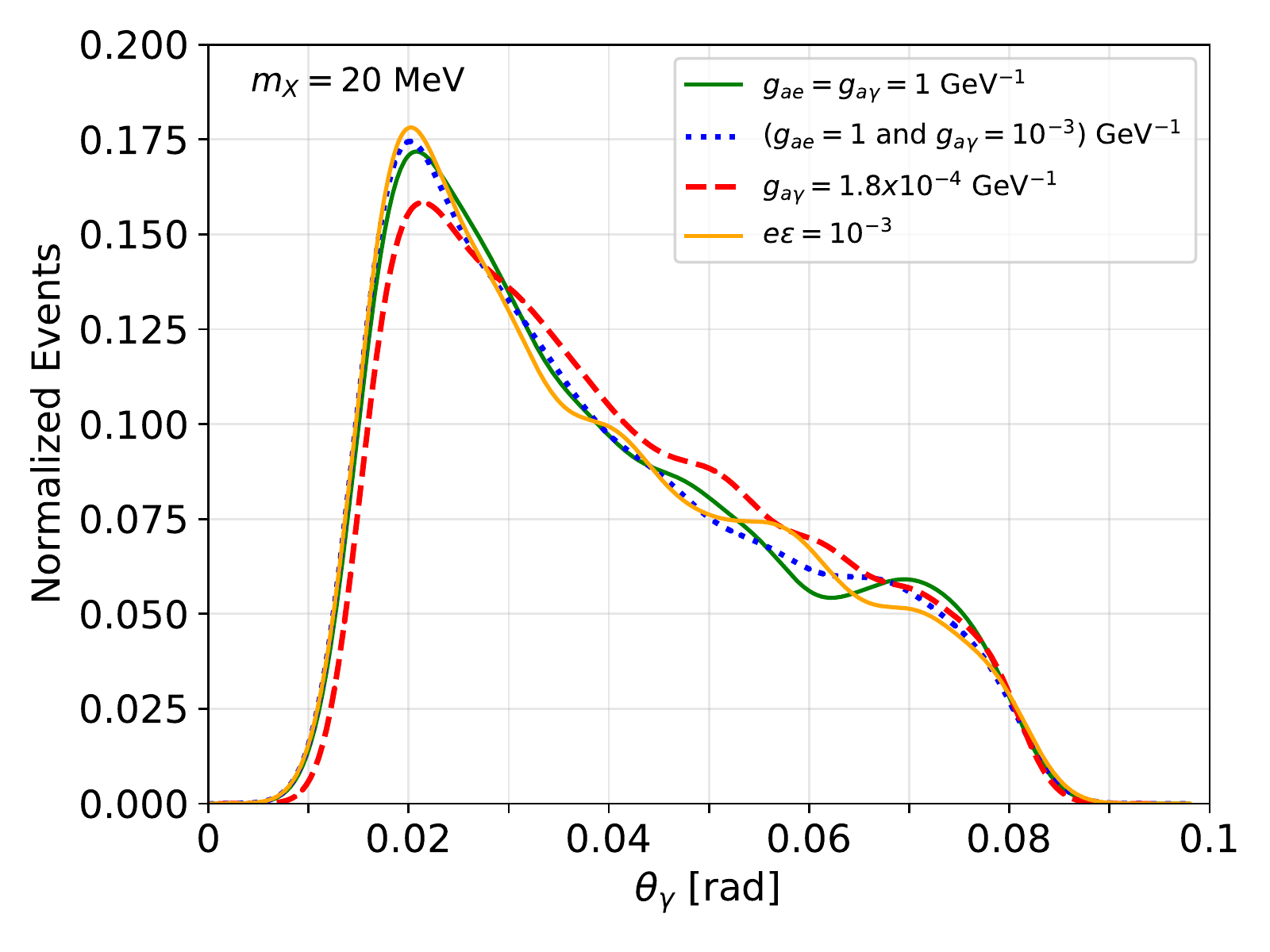}}
        \caption{Normalized distribution for the angle $\theta_\gamma$ 
        of a photon emitted in  $e^+e^-\rightarrow \gamma X$,  with $X$ a dark photon or an ALP, for $E_{beam}=550\,$MeV, and  $m_X=1\,$MeV (left), $m_X=10\,$MeV (right).
        For the ALP the couplings are  
        $\ge=\ga=1\,$GeV$^{-1}$ (green line), $\ge=10^{3}\ga= 1\,$GeV$^{-1}$ (blue dotted line) and 
        $\ge=0$, $\ga =1.8\cdot 10^{-4}\,$GeV$^{-1}$ (red dashed line).  
       For the dark photon (orange line) $e \epsilon=10^{-3}$.}
\label{fig:distrA}
\end{center}
\end{figure}
The missing mass resolution for an ALP candidate ($m_X=\ma$) 
depicted in in Fig.~(\ref{reso}) shows a good agreement with the 
PADME technical design report for a dark photon.
Including  selection cuts, the differential distribution for ALP events is 
--up to an overall normalisation factor--
roughly equivalent to the one for a dark photon (similarly to the case of BaBar).
 This is illustrated in Fig.~\ref{fig:distrA}, where we show the distribution of the photon emission angle w.r.t. to the beam direction $\theta_{\gamma}$, for the case of a dark photon and of an ALP (both for $s$-channel and associated production processes). We have 
 explicitly checked that also for PADME 
 the interference term between the $s$- and $t$-channel production processes 
  is negligible in the whole parameter range.

Altogether, we note that the higher ALP mass region is most affected by the cuts as the resonant behaviour of the associated (fermionic) channel, presenting a soft photon, is cut by the energy selection. This is illustrated

\begin{figure}[h!]
\begin{center}
\begin{subfigure}[b]{0.49\textwidth}
\includegraphics[width=\textwidth]{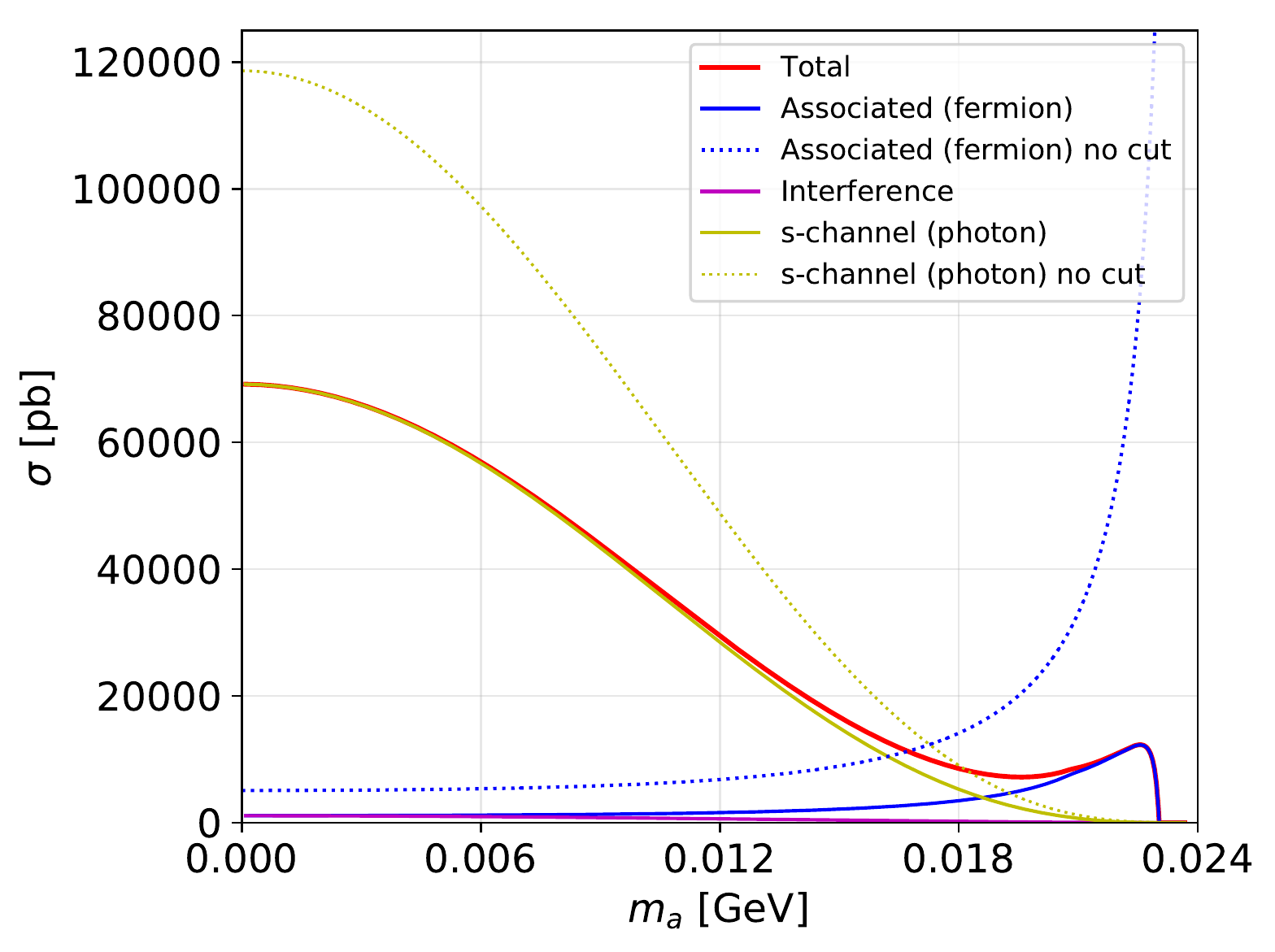}
\caption{ }
\label{fig:xsec11}
\end{subfigure}
\begin{subfigure}[b]{0.49\textwidth}
\includegraphics[width=\textwidth]{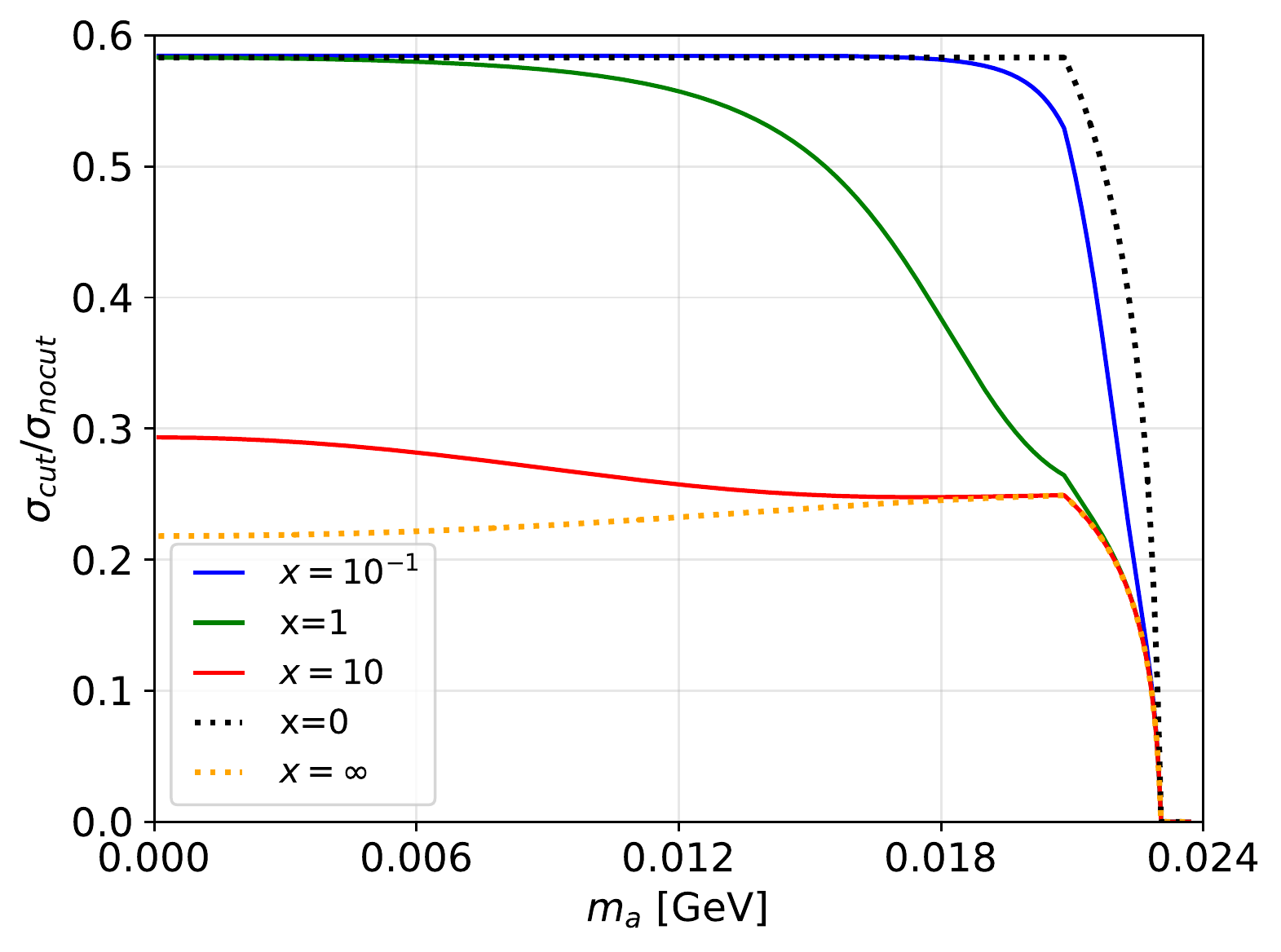}
\caption{ }
 \label{fig:Ratio}
\end{subfigure}
\end{center}
\caption{
(a) Cross-section for $e^+e^-\rightarrow \gamma  a$ as a function of $\ma$ for $\ge=\ga=1$ GeV$^{-1}$ and $E_{beam}=550$\,MeV with the  
PADME cuts $15\lesssim\theta_{\gamma}\lesssim 80\,$mrad, 30\,MeV $\lesssim E_{\gamma}\lesssim$\,500 MeV and without cuts.  
(b)~Selection efficiency 
for the total cross-section with the same cuts 
as a function of $\ma$ for different values of the  ratio $x=\ge/\ga$.  
}
\label{fig:PADMECS}
\end{figure}

In Fig.~\ref{fig:PADMECS}, in the first panel  we show the ALP production 
cross-section with and without cuts.
We see that the contribution of the $t$-channel  is heavily affected by the cuts 
in the large $\ma$ region. This is because the photon associated with 
the resonant regime is too soft and is cut out by the energy selection.
In the low mass range the photon channel dominates. We see that 
for this channel the cuts do not reduce the signal by more than 50\% in the whole 
mass range. 
In panel (b) we show the efficiency of the selection cuts for various 
values of the ratio $x=\ge/\ga$.
When the photon channel dominates over the fermion channel (dotted black line), the  
efficiency is about $60\%$ in all the mass range up to  $\ma \sim 21$ MeV. 
When only the fermion channel is present (orange line), 
the efficiency is much smaller, around $25\%$.
For $\ma\gtrsim 21$ MeV the energy cut reduces the signal down to 
zero, as in this case the photon energy is always $\lesssim 30$ MeV. 
From these results, it is clear that both annihilation
channels are important to search for the $e^+e^-\rightarrow 
a \gamma$ signal. On the other hand, as the current 
limits on the $\ga$ limits are more stringent than on 
$\ge$, PADME and its possible upgraded setups will be mostly 
relevant to probe the ALP coupling $\ge$.

The expected number of events at PADME can  be obtained from the total  cross-section as
\begin{align}
    N_{ALP}=N_{poT} \mathcal{N}_A \frac{Z_C \rho_C}{A_C} d_{t} \times (\varepsilon_{\rm PA} \sigma)_{\rm tot}
\end{align}
 $\rho_C = 3.5$g/cm$^3$ 
is the density of the diamond target, 
$A_C=12\,$g/mol, $Z_C = 6$, $d_{t}=100\,\mu$m 
is the target thickness, and $N_{poT}$ the number of positron on target. We have estimated the cross-section rescaled by the efficiency $(\varepsilon \sigma_{\rm tot})_{\rm PA} $ 
using \calchep and including the selection cuts for $\ge = 1$ GeV$^{-1}$ and $\ga = 1$ GeV$^{-1}$ independently, which we denote respectively as $(\varepsilon_{\rm PA} \sigma )_{e}$ and $(\varepsilon_{\rm PA} \sigma )_{\gamma}$. We thus have:
\begin{align}
   (\varepsilon_{\rm PA} \sigma)_{\rm tot} =\left( \frac{\ge}{1 ~\rm{GeV}^{-1}}  \right)^2 ( \varepsilon_{\rm PA} \sigma )_{e} + \left(\frac{\ga}{1 ~\rm{GeV}^{-1}}  \right)^2  (\varepsilon_{\rm PA} \sigma)_{\gamma}\,.
\end{align}
In order to estimate the background, we used the preliminary simulation described  in~\cite{Raggi:2015gza}. The original aim of the collaboration was to perform a `bump hunt' search in $M^2_{\rm miss}$. As a conservative estimate, we have used the background events below the smeared resonance peak at $1.5\sigma$ (corresponding to $~85\%$ of the signal). In estimating the future  reach with the POSEYDON setup, we have  
assumed that detector improvements combined with reduced pile-up would lead to roughly the same number of background events, despite the strong increase in statistics.
For the projection we will thus show a $100$-signal events line. Since the production rate depends on the square of the couplings, the projected limits can be rescaled to the desired number of signal events $n_s$ by multiplying by $\sqrt{100/n_s}$. 
To illustrate the maximum possible reach we further show the single event sensitivity (SES), corresponding to $2.3$ signal events. It is interesting to speculate that a possible 
improved PADME experiment running with the POSEYDON setup with $\sim 10^{16}$ poT/year, 
while inheriting the same detection technique, might rely on a different detector and   
hence on different background rejection power. The actual limit will therefore depend 
on the performance of the new detector.
In addition the forward region $\theta<$15 mrad as well as soft recoil photons, 
where most of the acceptance lies 
could be explored by a $10^{16}$ poT experiment, if the 
pile up could be reduced by factor of $\sim$ 100 with respect to the 
current PADME setup.

\begin{figure}[h]
\centering
        {\includegraphics[height=0.5\linewidth]{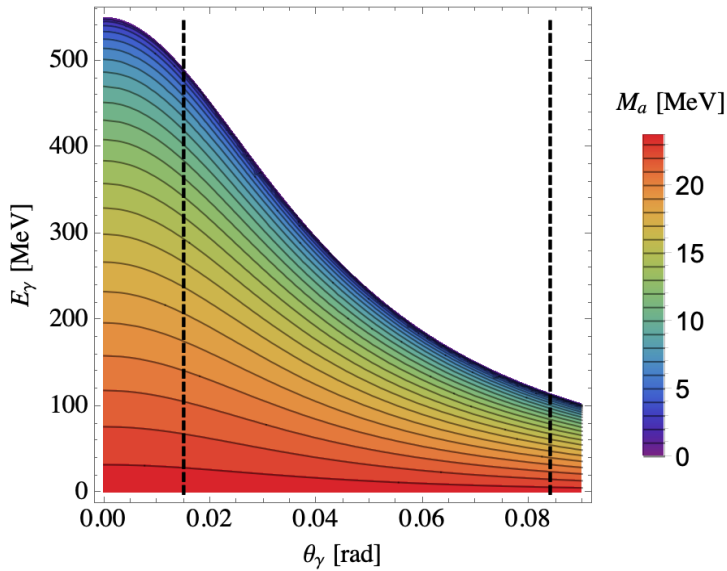}} \hspace{0.2cm}
        \caption{Correlation of the photon energy with its scattering angle at PADME. The gradient colour indicates the ALP mass. The dashed black lines are the angular separation between ECAL and SAC (Small Angle Calorimeter), and external ECAL border, respectively. The energy of the incident positron is $E_{beam}=550$ MeV.}
\label{fig:enthetacut550}
\end{figure}

Note that in case  a signal is observed, the correlation between the photon energy 
and the emission angle 
can provide information about the ALP mass. Fig.~\ref{fig:enthetacut550} shows how at small angles it is easier to distinguish different ALP masses, while at small $\ma$ 
the energy resolution of the calorimeter is crucial to pin down the precise value. 
The black dashed lines in the figure show the boundary of the ECAL region.

\subsection{Magnetic moment of light leptons}

The measured magnetic moment of the muon is one of the longest-standing anomalies in particle physics. Based on the original measurement at BNL~\cite{Bennett:2006fi} and on several improvements in the precision of the SM prediction (see e.g. the recent review~\cite{Aoyama:2020ynm}), the current discrepancy, parametrised in terms of $a_\mu ~\equiv~ \gmtwomu /2$ is:
\begin{align}
   \damu \equiv a^{\rm exp}_\mu  - a^{\rm SM}_\mu = (2.79 \pm 0.76) \cdot 10^{-9} \ ,
\end{align}
corresponding to a $3.7\sigma$ anomaly.  Intriguingly another anomaly 
was recently observed also in the anomalous magnetic moment of  the electron. 
Comparing the improved measurement of the fine structure constant from the study of cesium-133 atoms in 
a matter-wave interferometer~\cite{Parker:2018vye} with the SM theoretical prediction~\cite{Aoyama:2017uqe} yields:
\begin{align}
\label{eq:dae2018}
     \dae \equiv  a^{\rm exp}_e  - a^{\rm SM}_e = - (8.7 \pm 3.6) \cdot 10^{-13} \quad (\textrm{Berkeley-2018})\ ,
\end{align}
corresponding to a $2.4\sigma$ tension. Although the observation  of two discrepancies in the analogous  
observable for the muons and the electrons is indeed suggestive, 
the fact that the two deviations have discording signs poses a significant theoretical challenge
for finding an explanation in terms of new physics. 
Loop contributions from new light vector or scalar particles typically lead to a positive 
contribution to both the theoretical predictions for $a_e^{\rm th}$ and $a_\mu^{\rm th}$, while contributions from axial-vector or pseudo-scalar 
particles are instead negative. As 
the corrections are typically proportional to the square of the couplings of the new particles, 
one would expect that both deviations come with the  same sign.
\begin{figure}[]
\begin{center}
\begin{subfigure}[t]{0.3\textwidth}
\includegraphics[width=\textwidth]{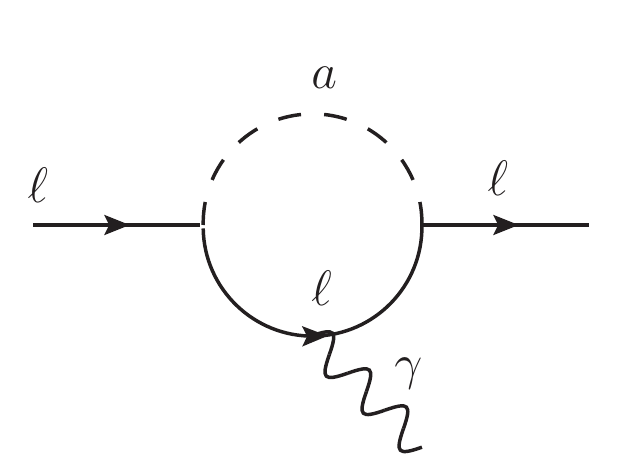}
\caption{}
\label{fig:gm2e}
\end{subfigure} 
\begin{subfigure}[t]{0.3\textwidth}
\includegraphics[width=\textwidth]{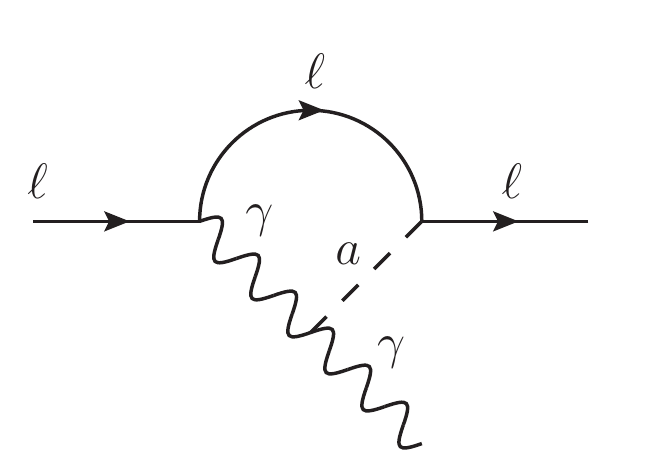}
\caption{}
\label{fig:gm2BarrZee}
\end{subfigure}
\end{center}
\caption{Feynman diagrams for the two main contributions of an ALP to $\dae$ and  $\damu$.  Diagram (\subref{fig:gm2e}) is proportional to $g_\ell^2$, while  the Barr-Zee diagram in  (\subref{fig:gm2BarrZee}) 
is proportional to  $\ga g_\ell$. }
\label{fig:diagsgmu}
\end{figure}
An ALP, that is a light pseudo-scalar, can provide one of the most elegant solution to explain both anomalies in one go, due to the simultaneous presence of the photon and the electron/muon coupling. The diagrams that yield the main contributions 
are drawn in  Fig.~\ref{fig:diagsgmu}. Crucially, the  second diagram in Fig.~(\subref{fig:gm2BarrZee}) is 
proportional to $\ga \gell$ and can therefore change sign depending on the sign of the ALP coupling 
to the particular fermion involved~\cite{Marciano:2016yhf,Leveille:1977rc}. Summing also the 
contribution of diagram (\subref{fig:gm2e})  (see, e.g.~\cite{Bauer:2019gfk}) as well as the  2-loop contribution 
from a light-by-light diagram~\cite{Marciano:2016yhf} leads to the overall correction:
\begin{align}
\label{eq:gm2}
     a^{\rm th}_\ell - a^{\rm SM}_\ell = \frac{m_\ell^2}{16 \pi^2} \left( 2 \ga \ge (h_2 - \log \frac{\Lambda^2}{m_\ell^2}) - \ge^2 h_1 +\ga^2 \frac{3 \aem }{\pi} \log^2 \frac{\Lambda}{m_\ell}\right)\ ,
\end{align}
where the loop functions $h_1(x)$ and $h_2(x)$, with   $x = {m_a^2}/{m_\ell^2}$, are:
\begin{align}
h_1(x) &= \int_0^1 dz \ \frac{2x^3}{x^2+z (1-x)} \\
h_2(x) &= 1 - \frac{x}{3} + \frac{x^2}{6} \log x + \frac{2 + x}{3}\sqrt{x(4 - x)}\arccos \frac{\sqrt{x}}{2} \ .
\end{align}
The possibility of a simultaneous fit to both anomalies on the basis of Eq.~\eqref{eq:gm2}
has been thoroughly explored in the literature.
The earlier attempts were based on a simplified model approach~\cite{Liu:2018xkx,Abu-Ajamieh:2018ciu}, 
and later lepton-flavour violating interactions were also used~\cite{Bauer:2019gfk,Cornella:2019uxs}. In the following, we  focus on the flavour-diagonal case, which can  explain both anomalies using only  Barr-Zee diagrams, assuming a 
GeV-scale ALP and the products of couplings  $\ga \ge \sim 10^{-5} \,\rm{GeV}^{-2}$ and $\ga g_{a \mu} \sim -10^{-6} \,\rm{GeV}^{-2}$.  

Several comments are in order. First, it is interesting to note that the contribution of $\ge$ to the electromagnetic anomaly (see Section~\ref{sec:model}), is precisely of the adequate order of magnitude to fit $\dae$ with $\ge \sim (0.01-0.1) \,\mathrm{GeV}^{-1}$. Additionally, while the required couplings to the fermions are of significant size,  
that this does not imply per-se that the corresponding scale $\Lambda$ has to be be very small. From a UV perspective, the ALP interactions arise proportionally to the charges of the  approximate broken global symmetry. A large coupling 
to a fermion species may thus simply indicate that this species boasts a large value of the global charge. 
The above reasoning has been  employed in various ways in 
the context of axion model building to boost selectively some axion couplings, see for instance~\cite{Darme:2020gyx}. We  also note that such a UV-complete theory with non-universal  ALP couplings to leptons will in general give rise  to flavor-violating ALP-lepton interactions. We refer to~\cite{Bjorkeroth:2018dzu,Cornella:2019uxs} for a thorough study of the constraints in this case. In the following, we will mostly focus on the case of the electron anomaly $\dae$, as we aim at constraining the ALP couplings to electrons and photons. We will nonetheless briefly comment also on the muon anomaly at the end of this work. As a final comment, it is clear that the contributions from Eq.~\eqref{eq:gm2} do not include possible direct effects from the UV theory.
Given that both anomalies are known to be strongly sensitive to the presence of new particles up to the TeV scale, it is advisable to consider these anomalies more as 
potential guidelines to identify some type of new physics, rather than using these measurements 
to constrain the ALP couplings  on and equal footing than the direct searches at accelerators.

Before concluding this section, it should be mentioned that the most recent atomic physics measurement of $\aem$
using Rubidium-87 atoms~\cite{Morel:2020} reported a value for $\aem$ that is more than $5\sigma$ away from the 
previous result of~\cite{Parker:2018vye} (and at $~\sim 2\sigma$ from an  earlier Rubidium-87 measurement). 
Taken at face value, this would imply 
\begin{align}
\label{eq:dae2020}
     \dae \equiv  a^{\rm exp}_e  - a^{\rm SM}_e = + (4.8 \pm 3.0) \cdot 10^{-13} \quad \rm{(LKB-2020)} \,,
\end{align}
still in mild tension with the SM but, most importantly, the deviation would now have a positive sign. 
Clearly, the $5\sigma$ discordance between the two most precise experimental measurements calls 
for a clarification,  and it would be highly desirable that an agreement on the values  of $\aem$
could be reached in the forthcoming coming years. As regards  our results, they can be straightforwardly adapted  
to either sign of the deviation, and this is because in this work  $\dae$ is the only observable which carries a 
dependence on the sign of $\ge$,  
and hence switching the sign of this coupling $\ge$ allows to fit both possibilities. In the following 
we will use the results from Eq.~\eqref{eq:dae2018}.\footnote{We 
have checked numerically that  the conclusions derived in the next section remain unchanged 
had we instead used Eq.~\eqref{eq:dae2020}. The only exception is that there is no 
parameter space region compatible with $\dae$ in the pure electrophilic case ($\ga=0$) shown in Fig.~\ref{fig:geonly}. 
However,  even the region compatible with Eq~\eqref{eq:dae2018} is already almost entirely excluded by other measurements.}

\section{Results}
\label{sec:results}

Relying on the analysis of  the previous sections, 
we are now in the position to discuss the updated status 
of searches for ALPs decaying (semi)-invisibly. 
In the plots of this section we will also underline when 
the relevant parameters fall in a range that could also explain 
the deviations from the SM predictions for the values of the   
$\mu$ and $e$  anomalous magnetic moments. 

\subsection{Photo-philic or electro-philic ALP}
\label{analbothcoupling1}

\begin{figure}[t]
\centering
        {\includegraphics[height=0.7\linewidth]{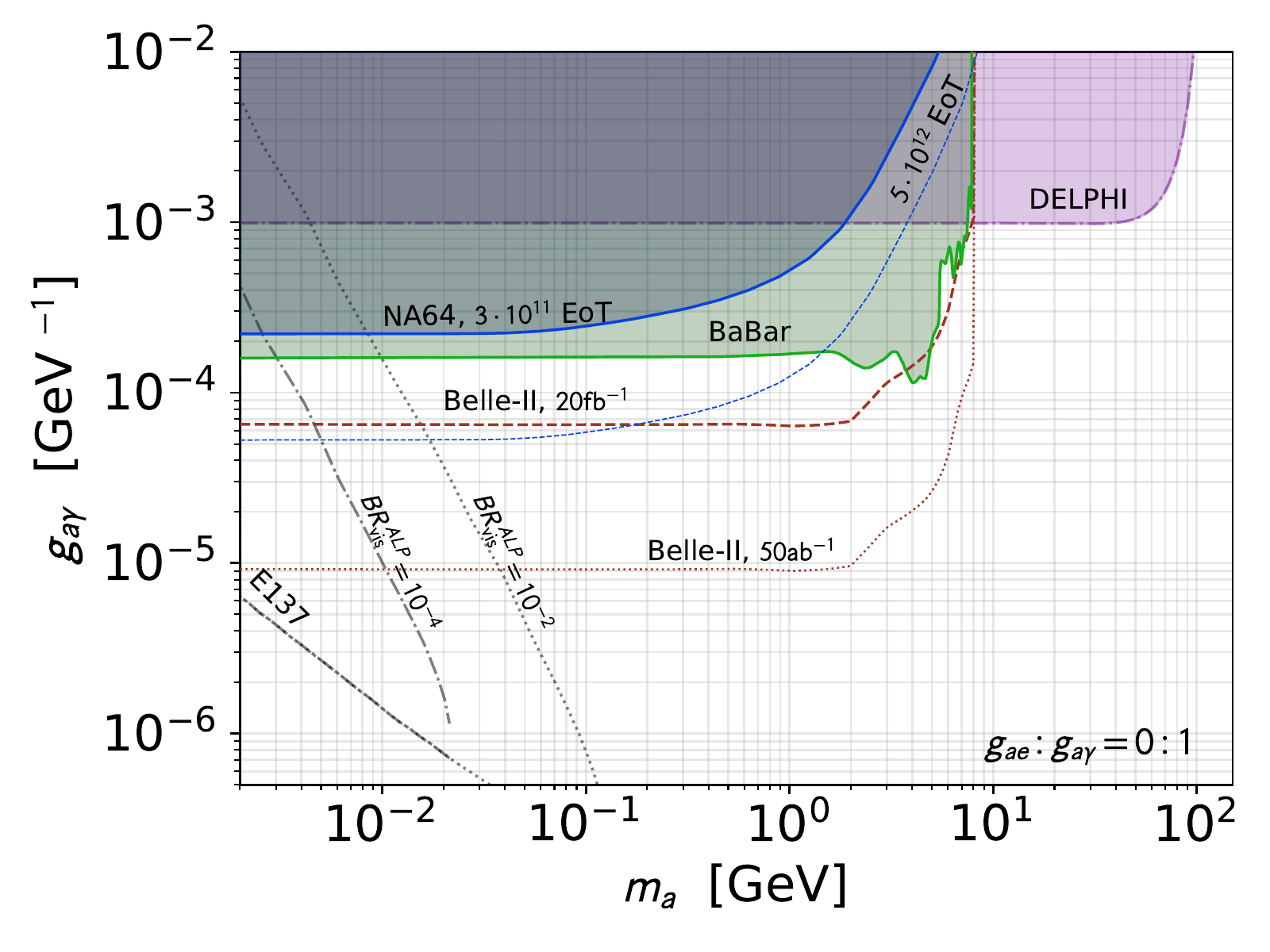}} \hspace{0.2cm}
        \caption{Summary of limits for an ALP decaying preeminently invisibly,  
        when $\ga$ dominates over $\ge$. The dotted and dash-dotted grey lines correspond to the range excluded by E137~\cite{Bjorken:1988as} for visible branching ratio respectively of  $10^{-2}$ and $10^{-4}$. The purple region is the exclusion from the DELPHI mono-photon search~\cite{Abdallah:2008aa,Fox:2011fx}. 
        The current (future) limits from NA64~\cite{Banerjee:2020fue} correspond to the solid (dotted) blue line. The green region is the exclusion from mono-photon search at BaBar~\cite{Lees:2017lec}. The dashed (dotted) rust line represent the projected reach of Belle-II~\cite{Kou:2018nap} for a $20\,\rm{fb}^{-1}$ ($50 \,\rm{ab}^{-1}$) dataset.}
\label{fig:resga}
\end{figure}

Let us first consider the  case of a photo-philic ALP, that is an ALP  for which  the coupling 
$\ga$ is the dominant one.  The corresponding limits are presented in Fig.~\ref{fig:resga}. 
From this plot it is clear that the mono-photon search from BaBar~\cite{Lees:2017lec} dominates the constraints 
in most of the parameter space. In particular, and in contrast with the case of a dark photon or of a visibly decaying ALP, the NA64 missing energy search~\cite{Banerjee:2020fue} is so far providing subdominant constraints.
Nonetheless, with the full $5\cdot 10^{12}$ EoT dataset, NA64 could eventually take over BaBar,
and compete with the Belle-II reach based on a first dataset of $20\,$fb$^{-1}$. 
In the long run, the results from Belle-II will likely dominate the limits by about an order of magnitude. 
Limits from  PADME do not lead to new exclusion regions and are not shown in the plot.

Interestingly, the limits from visible searches in beam dump experiments, such as E137, can remain relevant even when 
the ALP visible decay branching ratio is largely suppressed. 
 This is shown in Fig.~\ref{fig:resga}: even with a visible branching ratio $\rm{BR}_{\rm vis}^{\rm ALP} \equiv (\Gagam + \Gae) / \Gamma_{inv}$ as small as $10^{-4}$ (dot-dashed grey line), the E137 limit remains relevant in the 
 low mass region, due to the fact that in this region the BaBar and NA64 limits saturate. 
 This shows that for ALPs decaying partially invisibly, all beam dump limits need to be estimated carefully. 
 Note that we do not show the limit from $\dae$, since the strong constraints on the photon coupling
 imply that the light-by-light contribution from Eq.~\eqref{eq:gm2} is  negligible
 (besides having the wrong sign to accommodate the  measurement of $\dae$).

\begin{figure}[t]
\centering
        {\includegraphics[height=0.7\linewidth]{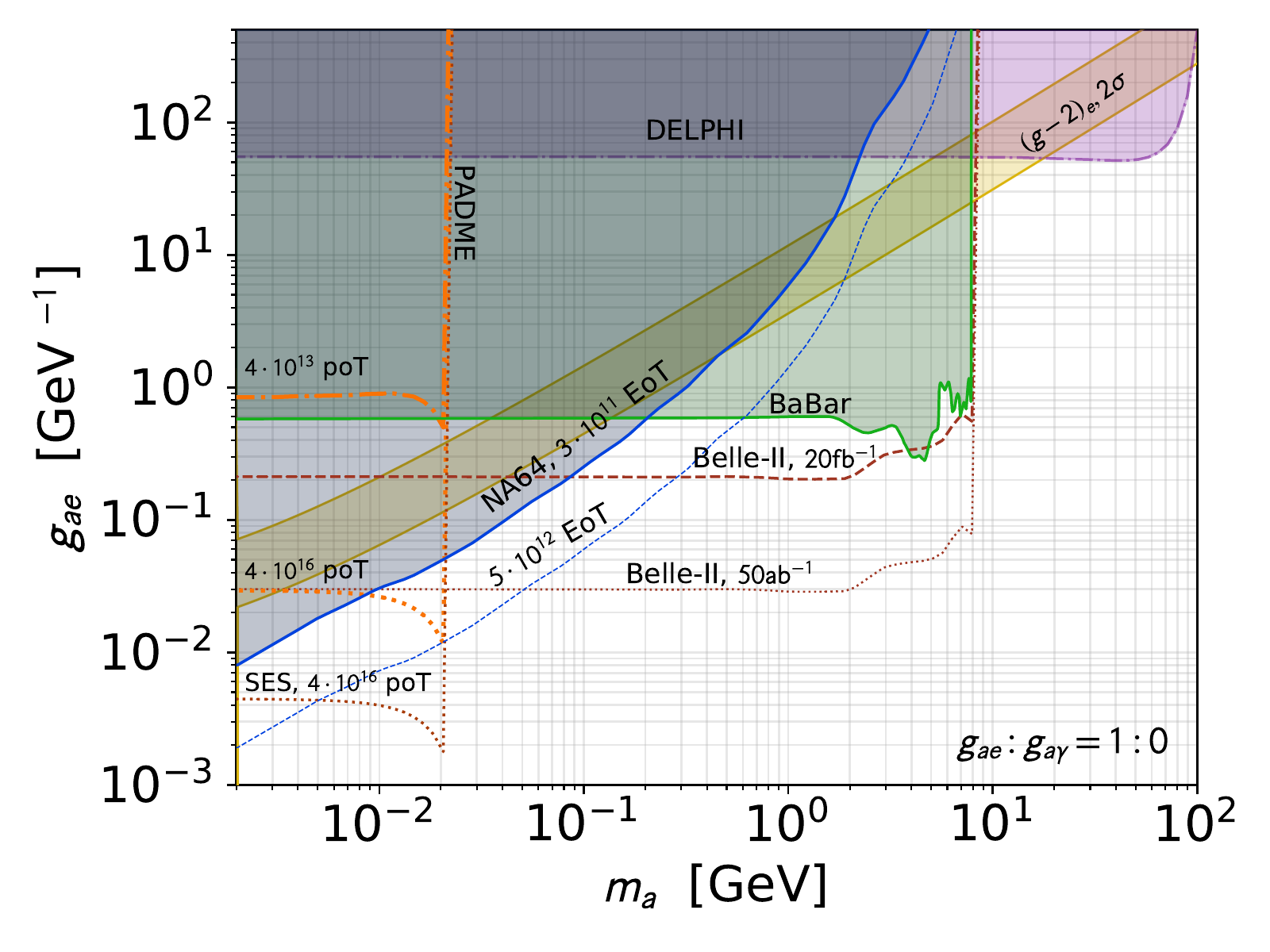}} \hspace{0.2cm}
        \caption{Summary of the limits when  $\ge$ dominates over $\ga$. The three broken orange lines are 
        (top to bottom) the projection for PADME~\cite{Raggi:2015gza} for the current run, and for a POSEYDON~\cite{Valente:2017hjt} upgrade of the beam with either $100$ signal events, or with 
        single-event sensitivity (SES - $2.3$ events line). The purple region is the exclusion from DELPHI mono-photon searches~\cite{Abdallah:2008aa,Fox:2011fx}. The solid (dotted) blue line corresponds to  the current (future) limit from NA64~\cite{Banerjee:2020fue}. The green region is the exclusion from mono-photon searches at BaBar~\cite{Lees:2017lec}. The dashed (dotted) rust line gives the prospects at Belle-II~\cite{Kou:2018nap} for a $20 \,\rm{fb}^{-1}$ ($50 \,\rm{ab}^{-1}$) dataset. The golden strip corresponds to the $2\sigma$ range for 
        the $\gmtwoe$ anomaly.}
\label{fig:geonly}
\end{figure}

In Fig.~\ref{fig:geonly} we present 
the case of a purely electrophilic ALP ($\ga=0$).
Due to the scaling of the electron bremsstrahlung cross-section with the electron-ALP mass ratio 
(see Eq.~\eqref{eq:bremvsPrim}) the limits from electron beam dump, and in particular from NA64, 
get stronger with  decreasing $\ma$. In line with the general scaling argument presented in 
Sec.~\ref{sec:model}, the limits follow the same hierarchy as for the better known case of a dark photon portal. 
NA64 dominates the limits in the lower mass region (below around $200$ MeV), while BaBar and soon-to-be release results from Belle-II provide the strongest constraints up to around $10$ GeV. Probing  higher ALP masses is significantly harder for  intensity frontier experiments, so that LEP mono-photon searches still provide the strongest bounds in the tens of GeV range. It is worth noting that in this region the `pure' $\ge$ contribution to $\dae$ could still accommodate the experimental measurement, although most of the useful  parameter space is already excluded.  With an upgrade of the   
experimental setup, in a five-year time-scale PADME  could probe  part of the  parameter space accessible 
at NA64 only after  the  full expected dataset will be collected.  

\subsection{Combined case}
\label{analbothcoupling3}

\begin{figure}[h!]
\begin{center}
\begin{tabular}{c}
\begin{subfigure}[b]{0.6\textwidth}
\includegraphics[width=\textwidth]{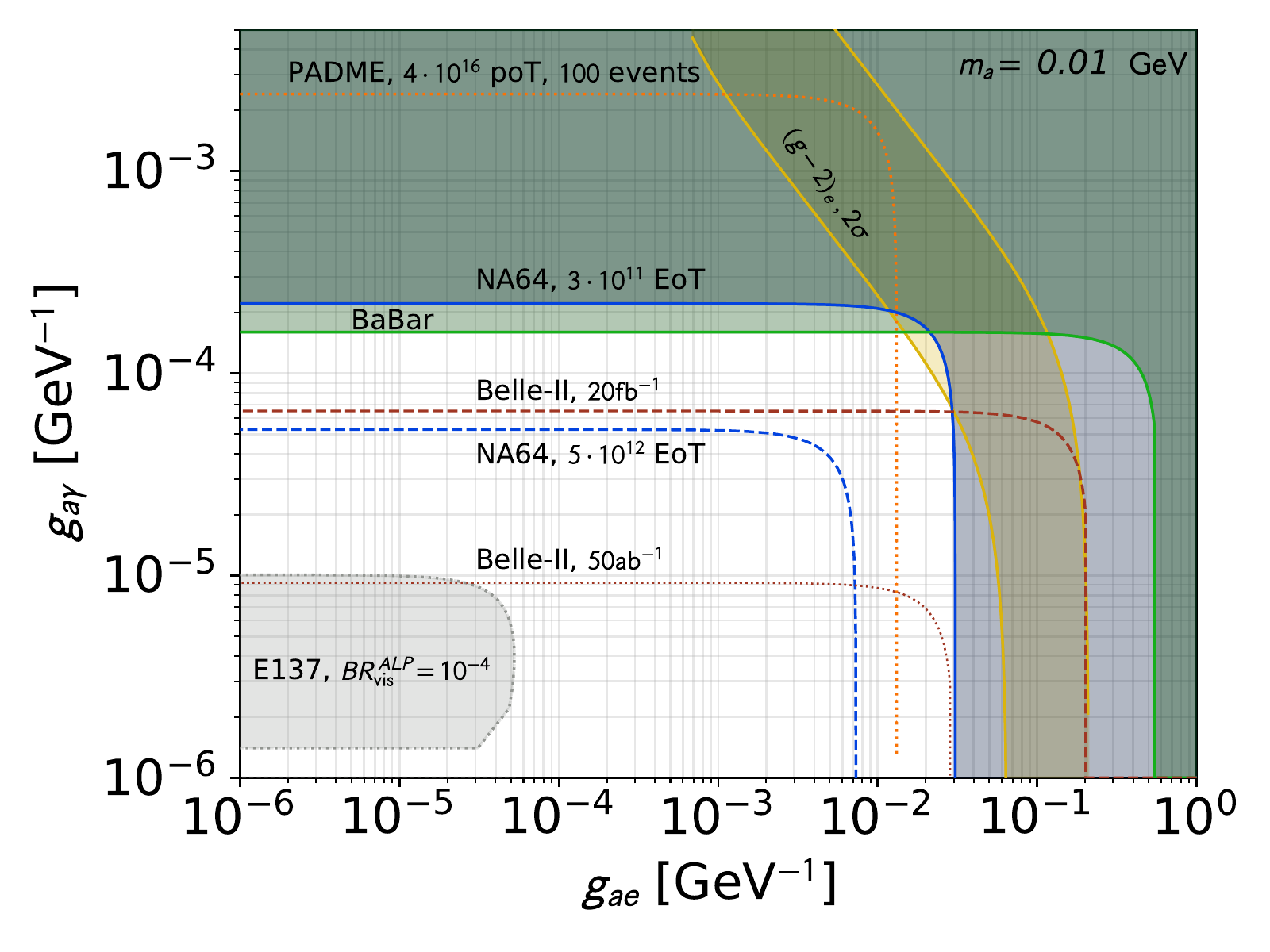}
\caption{ }
\label{fig:ma001}
\end{subfigure}
\vspace{-0.1cm} 
\end{tabular}\\
\begin{tabular}{cc}
\begin{subfigure}[b]{0.49\textwidth}
\includegraphics[width=\textwidth]{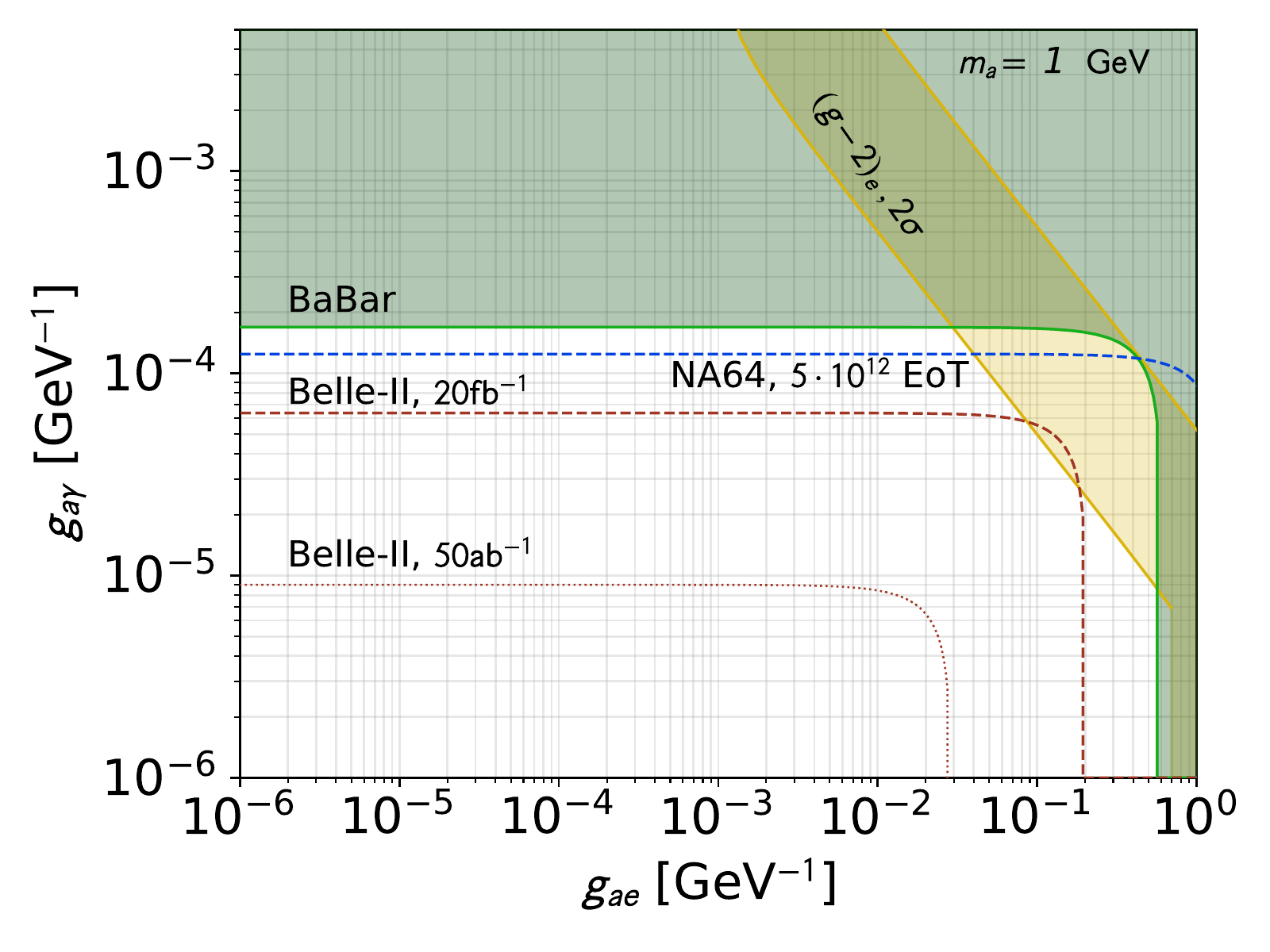}
\caption{ }
 \label{fig:ma1}
\end{subfigure} &\begin{subfigure}[b]{0.49\textwidth}
\includegraphics[width=\textwidth]{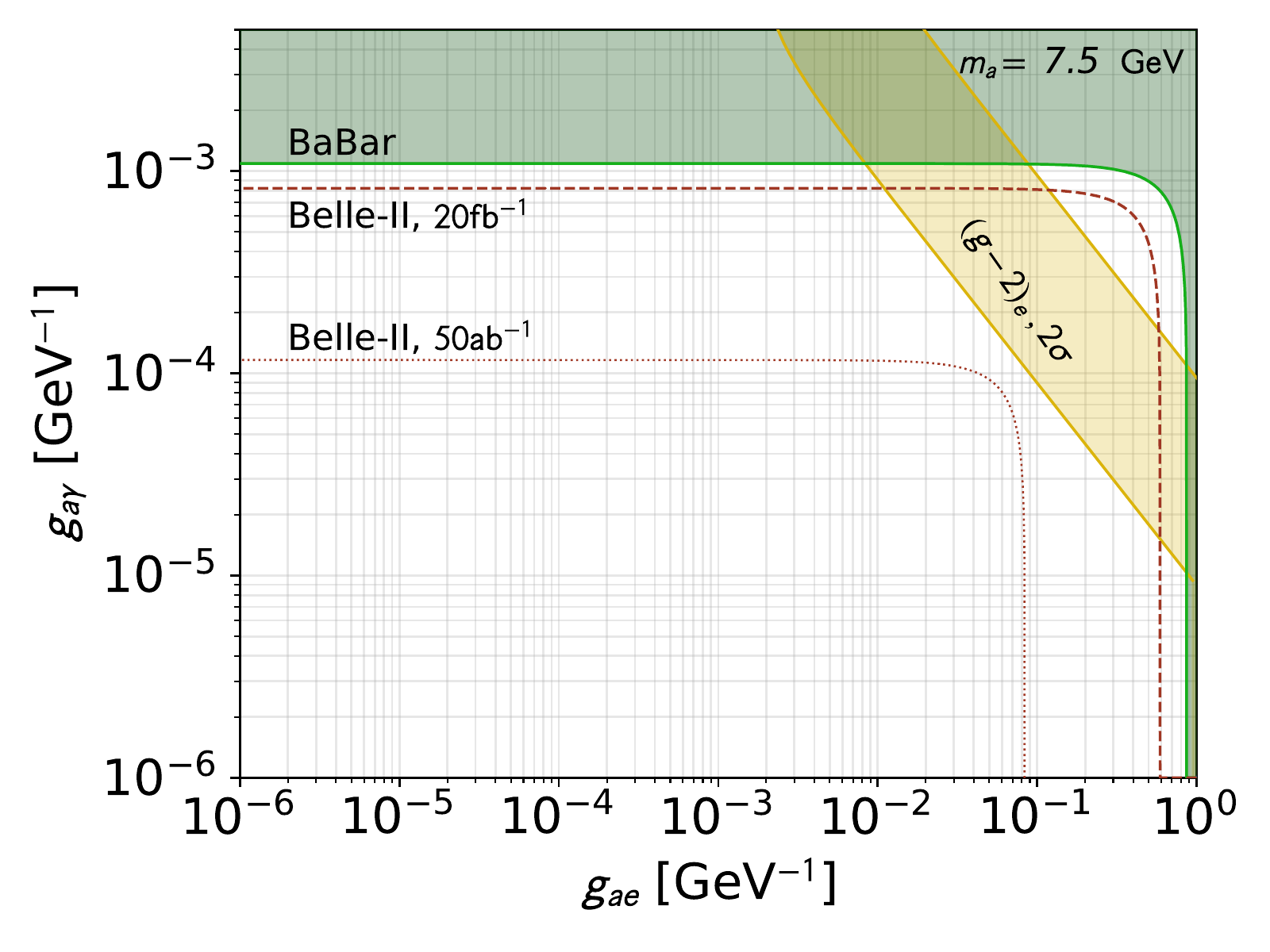}
\caption{ }
\label{fig:ma75}
\end{subfigure}
\end{tabular}
\vspace{-0.5cm} 
\end{center}
\caption{Summary of limits as function of $\ge$ and $\ga$ for $m_a= 10$ MeV (\ref{fig:ma001}), $m_a= 1$ GeV (\ref{fig:ma1}) and $m_a= 7.5$ GeV (\ref{fig:ma75}). The grey region in \ref{fig:ma001}) is excluded 
by  E137~\cite{Bjorken:1988as} assuming a  visible branching ratio of $10^{-4}$. The dashed orange 
lines is the projection for a POSEYDON~\cite{Valente:2017hjt} extension of PADME with $100$ signal events. The purple region is the exclusion from the DELPHI mono-photon search~\cite{Abdallah:2008aa,Fox:2011fx}. 
The current (future) limits from NA64~\cite{Banerjee:2020fue}
corresponds to the solid (dashed) blue line.
The green region is the exclusion from mono-photon search at  BaBar~\cite{Lees:2017lec}. The dashed (dotted) rust line represents the prospects at Belle-II~\cite{Kou:2018nap} for a $20 \,\rm{fb}^{-1}$ ($50 \,\rm{ab}^{-1}$) dataset. 
The golden strip corresponds to the $2\sigma$ range for the $\gmtwoe$ anomaly.}
\label{fig:combinedlim}
\end{figure}
Let us now turn to the case where both couplings are present and they are both relevant. 
We show the limits for this case in the three panels of Fig.~\ref{fig:combinedlim}, 
for  $\ma = 0.01,\,1$ and $7.5$ GeV. In most cases the presence of both couplings does 
not lead to a striking new behaviour. For instance, for NA64, PADME, BaBar and 
Belle-II (projective) limits, a smooth transition is observed when both 
the electron and photon contributions to the ALP production rates are of similar size. 
In the case of the E137 limits, note that although we have only included ALP production via the Primakoff mechanism, 
the electron channel is the main visible ALP decay mode in the right-hand part of 
the three plots. 
A remark is in order for the surviving  limit that can be obtained from 
beam dump experiments for $\ma$ in the few MeV range.   This is illustrated 
by the E137 excluded region shown as the grey area in Fig.~\ref{fig:ma001}, which 
corresponds to an ALP visible branching ratio of $10^{-4}$.
The assumption that the ALP dominantly decays invisibly, implies that 
for $\ma = 10$ MeV and $\ge \gtrsim 5\cdot 10^{-5}\,$GeV$^{-1}$ the 
invisible decay length becomes much smaller than the length of the 
shielding, so that there is no bound beyond this point. 
The second observable which depends non-trivially on both couplings is the 
correction to the electron anomalous magnetic moment $\dae$. The combination of the 
pure $\ge^2$ contribution  and of the  Barr-Zee $\ge \ga$ term  opens up  
an interesting region in parameter space, that remains below the current NA64 and BaBar limits,
for ALP masses larger than about $10\,\mathrm{MeV}$. 
The experimental prospects to probe this region are rather good, hence the $(g-2)_e$  anomaly 
constitutes an obvious target for future improved limits from NA64, PADME and  Belle-II. 
The $2\sigma$ region wherein the discrepancy can be explained  
(shown as the gold strip in the three figures) could  be likely  covered almost entirely, 
in particular thanks to  future Belle-II limits. 

Finally, the interplay between the limits from NA64 and BaBar is of interest. 
The former is relevant only for low ALP masses, while the latter  provides significant constraints 
also for $\ma \gtrsim 1\,$GeV. Moreover, it can be seen from 
Fig.~\ref{fig:ma001} that for $\ma \sim 10\,$MeV   BaBar  dominates the limits 
on $\ga$, while   NA64 provides the strongest constraints on $\ge$. 
This can be traced back in  large part to the $1/s$ suppression 
of the associated cross-section in electron-positron colliders.

\begin{figure}[t]
\begin{center}
\begin{subfigure}[b]{0.49\textwidth}
\includegraphics[width=\textwidth]{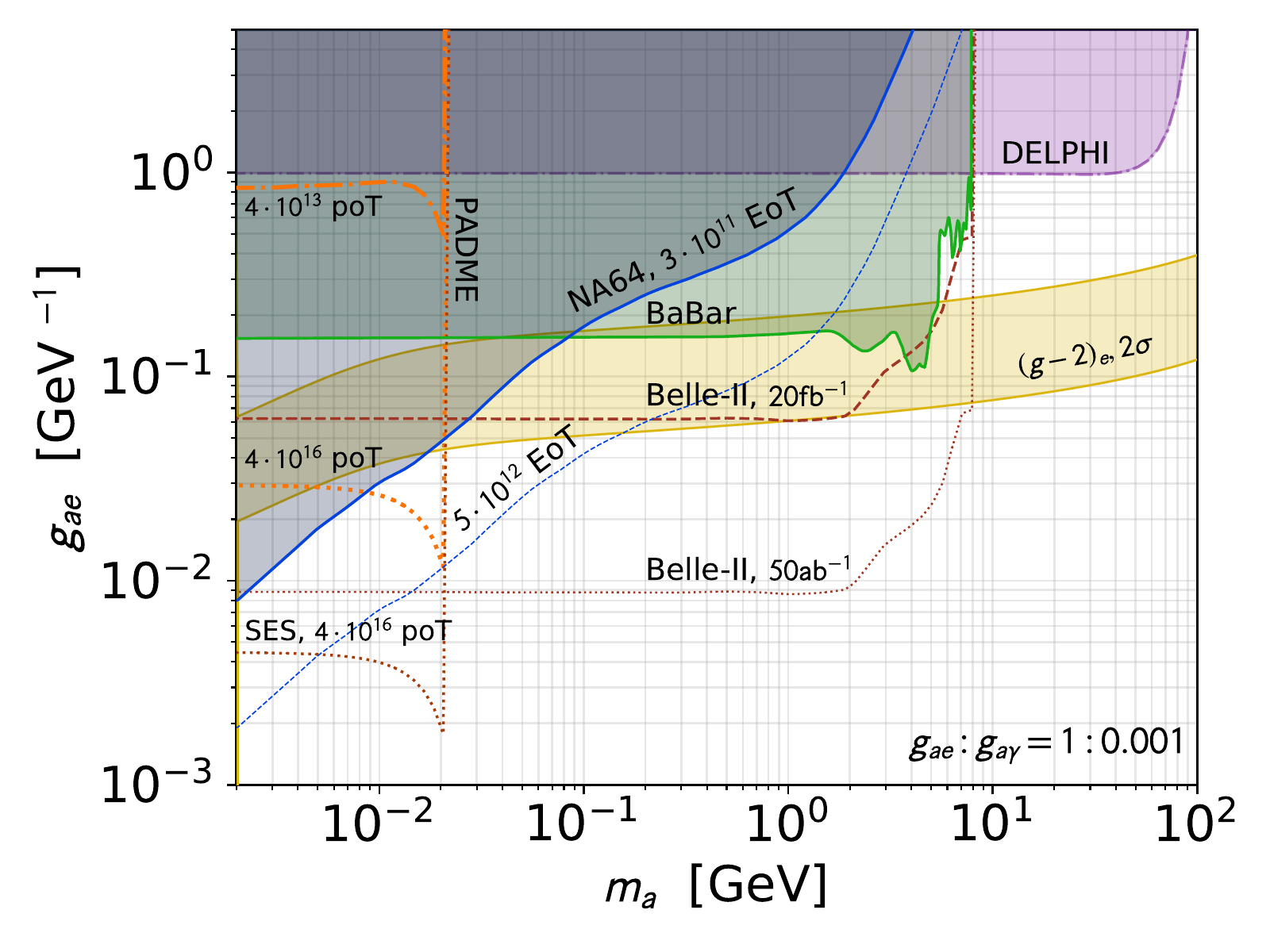}
\caption{ }
\label{fig:gage1}
\end{subfigure}
\begin{subfigure}[b]{0.49\textwidth}
\includegraphics[width=\textwidth]{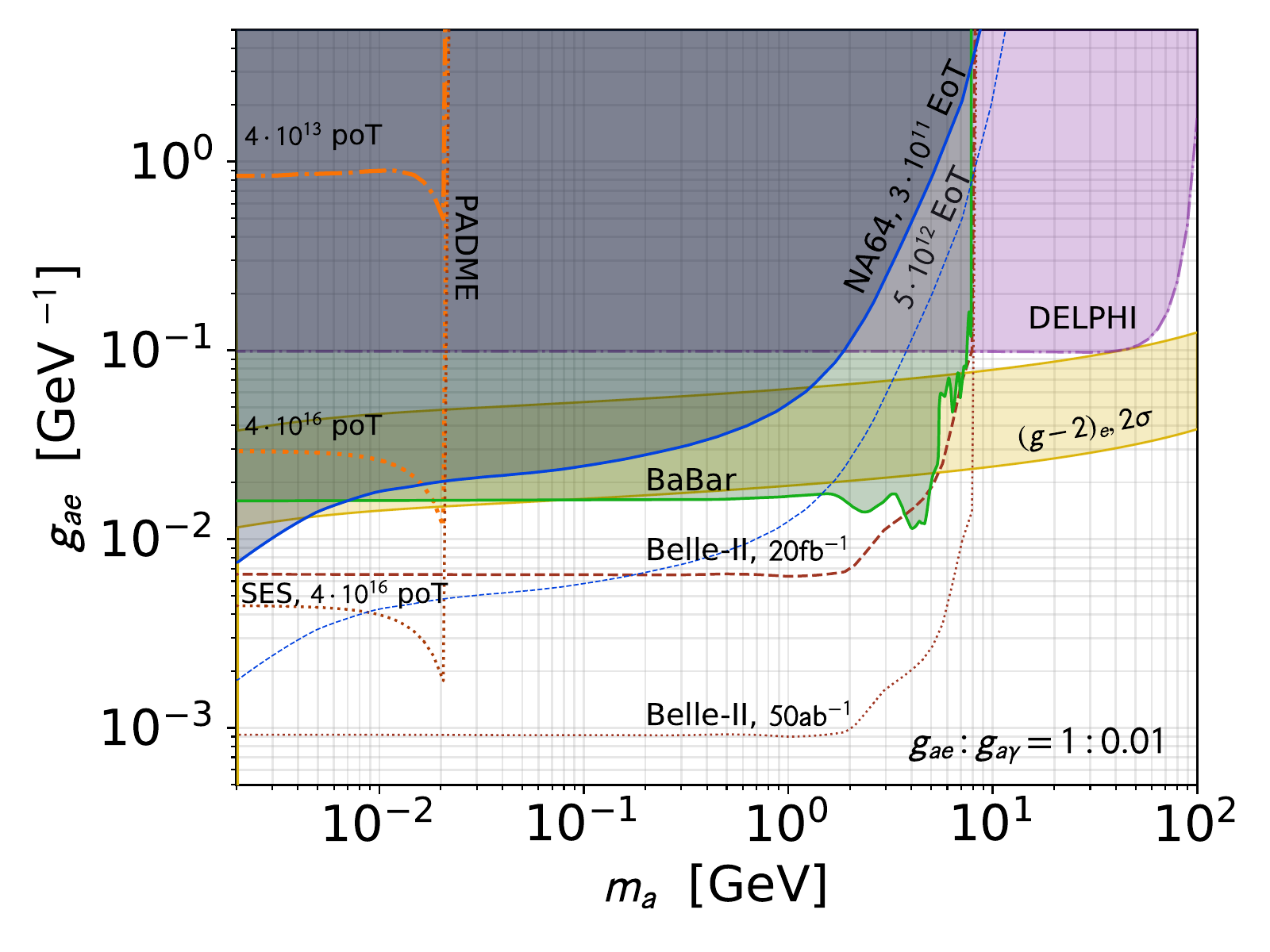}
\caption{ }
 \label{fig:gage2}
\end{subfigure} 
\vspace{-0.2cm} 
\end{center}
\caption{Summary of limits as function of $\ge$ and $\ma$ for $\ge/\ga= 10^3$ (\ref{fig:gage1}) and 
$\ge/\ga = 10^2$ (\ref{fig:gage2}). A negligible ALP visible branching ratio is assumed. The dashed orange lines are (top to bottom) the projection for PADME~\cite{Raggi:2015gza} for the current run, and for a POSEYDON~\cite{Valente:2017hjt} extension with either $100$ signal events, or with single-event sensitivity (SES - $2.3$ events line). The purple region is the exclusion from DELPHI mono-photon searches~\cite{Abdallah:2008aa,Fox:2011fx}. 
The current (future) limits from NA64~\cite{Banerjee:2020fue} correspond to the solid (dotted) blue line.
The green region is the exclusion from BaBar~\cite{Lees:2017lec}. The dashed (dotted) rust line 
gives the projection for Belle-II~\cite{Kou:2018nap} for a $20 \,\rm{fb}^{-1}$ ($50 \,\rm{ab}^{-1}$) dataset. 
The golden strip corresponds to the $2\sigma$ range for the $\gmtwoe$ anomaly.}
\label{fig:combinedlimmass}
\end{figure}

\begin{figure}[t]
\centering
        {\includegraphics[height=0.6\linewidth]{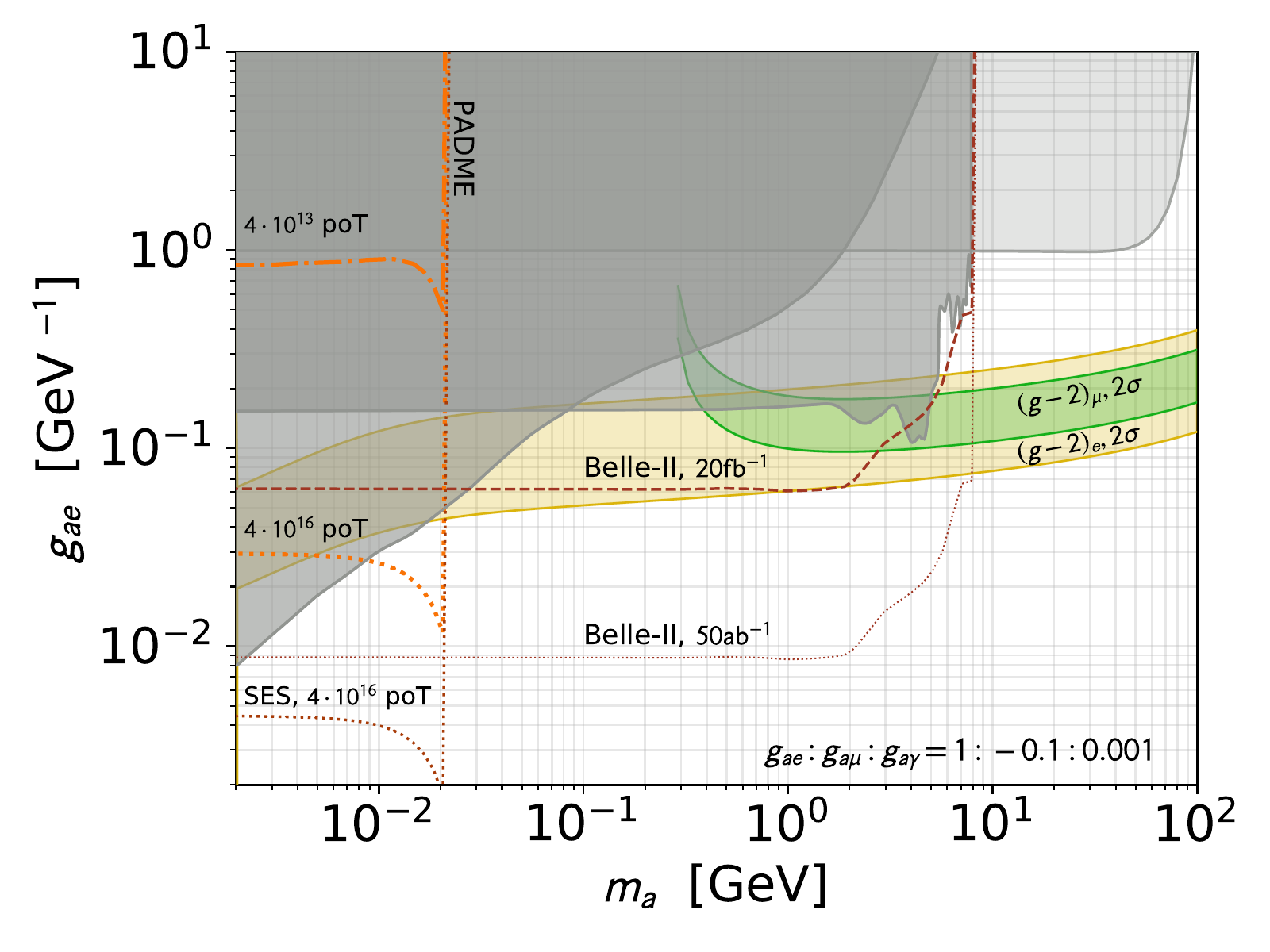}}
        \caption{The invisible ALP $(g-2)_{e, \mu}$ solution. Summary of limits for $\ge/\ga =10^3$ 
        and $g_{a\mu}/\ga = -10^2$. The dashed orange lines  (from top to bottom) give the projection for PADME~\cite{Raggi:2015gza} for the current run, and for a POSEYDON~\cite{Valente:2017hjt} extension with either $100$ signal events, or with single-event sensitivity (SES - $2.3$ events line). The grey regions are exclude by DELPHI~\cite{Abdallah:2008aa,Fox:2011fx}, NA64~\cite{Banerjee:2020fue} and BaBar~\cite{Lees:2017lec}. The  dashed (dotted) rust line 
        shows the projected limits from Belle-II~\cite{Kou:2018nap} for a $20 \,\rm{fb}^{-1}$ ($50 \,\rm{ab}^{-1}$) dataset. The golden strip corresponds to the $2\sigma$ range for
        explaining the $\gmtwoe$ anomaly, and the green region to the $2\sigma$ range for 
        the $\gmtwomu$ anomaly.}
\label{fig:gm2}
\end{figure}

We investigate further the mass dependence of these limits in Fig.~\ref{fig:combinedlimmass}. We show the limits on the ALP couplings as function of $\ma$  for $\ge/\ga = 10^3$ in 
Fig.~(\ref{fig:gage1}) and for $\ge/\ga = 10^2$ in Fig.~(\ref{fig:gage1}). The first choice corresponds roughly to the expected size of $\ga$ when is 
generated radiatively from an electron loop. In both cases, the limits are typically dominated by the photon-mediated processes, particularly for BaBar and Belle-II, while for NA64  
for large values of  $\ma$ the limits are determined by $\ga$, but become  
dominated by $\ge$ as the ALP mass is decreased. 
The figure also illustrate the large parameter space available for an electrophilic ALP
($\ge\gg \ga$) solution to the $\dae$ anomaly, with the viable parameter space extending down 
to tens of MeV. In the absence of any significant compensating UV contribution, one can regard 
the area above the golden strip as excluded by the $\dae$ measurement. However, by construction 
the ALP Lagrangian Eq.~\eqref{eq:lagr} defines a low energy effective theory, which 
assumes the presence of an UV sector with a larger field content. One should 
therefore be cautious regarding limits derived from $(g-2)_e$ because additional  
contributions to from the UV sector to this quantity are certainly possible.  

We conclude this section by extending the ALP interactions to include also 
an interaction with the muon controlled by the coupling  $g_{a\mu}$.
This renders possible to fit simultaneously both the $\dae$ and $\damu$ anomalies.  
Fig.~\ref{fig:gm2} shows the preferred region in  
the $(m_a,\ge)$ plane for a ratio of the fermion couplings $\ge/g_{a\mu} \sim - 10 $.
As expected from previous results (see e.g.~\cite{Cornella:2019uxs}) the choice 
of this value for the ratio of the lepton couplings 
provides a very good fit in the range of large ALP masses. However, we find   
that the combined region extends to mass values much lower than what was previously though, 
as the 
BaBar results are in fact not sufficient to exclude the $\damu$-favoured 
region below $10$ GeV. Moreover, it is interesting to remark that the 
prediction from a radiatively generated ALP-photon coupling  $\ga/\ge \sim 10^{-3}$ 
leads to an adequate range for the photon interaction. Indeed this 
ratio offers a good compromise between a suppression of accelerator and collider limits 
(in particular from BaBar) and a sufficiently large enhancement of both $\dae$ and $\damu$ via 
the Barr-Zee contribution. Note that we have  not included  limits 
on $g_{a\mu}$ from  the recast of past beam dump experiments 
(e.g. E137 limits~\cite{Marsicano:2018vin}) or projected limits from  
possible future experiments such as  NA64-$\mu$~\cite{Chen:2018vkr}, $M^3$~\cite{Kahn:2018cqs} 
or from dedicated Belle-II searches~\cite{Adachi:2019otg}. This is because the fit requires 
an  ALP muon coupling sizeably smaller than the electron one, 
hence we expect that future limits on $\ge$ will keep providing the strongest 
constraints for the ALP solution to the $g-2)_{e,\mu}$ anomalies.

\section{Conclusions}
\label{sec:conclusions}

In this work we have presented  a thorough analysis of the experimental 
constraints on the scenario in which an ALP constitutes a portal to a dark 
sector, and  dominantly decays into light dark particles, while in the SM sector 
it couples to photons and electrons to which it can also decay, but 
with subdominant branching ratios. 
We have first reviewed existing astrophysical constraints on this scenario, 
questioning in particular the limits from the duration of the SN1987 neutrino burst, 
which rely on the fact that the burst duration would be shortened 
if the ALP decay products could freely escape from the proto-neutron star core.
We have in fact shown that interactions within the dark sector of a reasonable strength 
may  suffice to keep the light dark particles trapped inside the core, along with the ALPs, 
on the same time scale of neutrino trapping. 
We have then proceeded to perform a thorough re-interpretation of all the relevant collider and 
beam dump experiments relevant to constrain this scenario. 
Such a study was not yet available for the case of an ALP 
decaying dominantly into invisible channels. 
Not surprisingly, we have shown that the stronger constraints arise from various missing 
energy searches in BaBar and NA64, and from and mono-photon searches at LEP, and in particular 
from the results reported by the DELPHI collaboration~\cite{Abdallah:2008aa}. 
Although the DELPHI mono-photon search  results have been already widely used to 
constraint both an invisible dark photon and an invisible ALP, we have 
accounted for the first time for the variation of the LEP center-of-mass energy during 
the run time. This has the important effect of shifting the mass threshold of the analysis 
to larger ALP masses. We have reinterpreted  NA64 limits on visible ALP decays 
in terms of limits on invisible decay channels, mimicking to a certain extend the 
results found by the collaboration for the case of an invisible dark photon. 
We have also shown that  relaxing completely the limits from existing searches 
for long-lived ALPs is challenging,  and requires a significant suppression of the 
visible branching ratios. We have illustrated this point for the particular case of the SLAC 
E137 beam dump experiment, however,  the procedure that we have described allows in principle 
to recast in the same way the results of other beam-dump based searches for long-lived ALPs. 
Finally we have investigated the reach of the current run of the PADME positron beam dump experiment, 
and we have discussed the projected sensitivity  in case the positron beam delivered 
by the LNF Beam Test Facility could be upgraded according to the POSEYDON proposal.

An invisible ALP as the one investigated in this work is among the  
preferred new physics candidates for explaining the discrepancy between 
the predicted value of the anomalous magnetic moment of the electron and the 
result of a recent measurement. 
We have confronted the parameter range hinted at by this discrepancy with the limits 
we have derived on the ALP mass and couplings, 
finding that a sizeable parameter space region  is still viable, which thus constitutes 
an important target for upcoming experiments. 
In particular, we have argued that both the NA64 and  Belle-II experiments will be able to  
probe in the  near future a large part of this region for an ALP mass below $10\,$GeV. 
We have additionally discussed the possibility of a simultaneous explanation 
for  the deviations of the $(g-2)_e$ and $(g-2)_\mu$  measured values from the SM predictions. 
We have shown that if the ALP couples to the muon with a coupling about a factor of ten  
smaller than the electron coupling, and of  opposite sign, the ALP explanation is viable  
for ALP masses as low as a few hundreds of MeV.  
This observation provides an additional golden target for future experimental studies.

\paragraph{Note added:} Shortly after this paper first appeared, 
the result of a new experimental measurement of the muon anomalous magnetic moment 
by the FNAL Muon g-2 experiment was published~\cite{Abi:2021gix}, that  when 
combined with the old BNL result~\cite{Bennett:2006fi} yields the new world 
average 
\begin{align}
   \damu \equiv a^{\rm exp}_\mu  - a^{\rm SM}_\mu = (2.51 \pm 0.59 ) \cdot 10^{-9} \ .
\end{align}
While this new measurement increases the tension with the SM prediction at the level of 4.2$\sigma$,
it does not significantly modify the central value. Thus 
our results, and in particular Fig.~\ref{fig:gm2},  remain compatible with the new 
experimental value.

\subsection*{Acknowledgments}
\noindent
We acknowledge discussions with L.~Delle Rose and R.~Simeonov at the beginning of the project, P.~Valente for providing us with the details of the DA$\Phi$NE POSEDYON setup, and G. Corcella for important advice on the numerical aspects of this work.  
We thank P.~Paradisi for useful comments on the first version of the paper, 
and M.~Hirsch for bringing to our attention Ref.~\cite{Morel:2020}. 
L.D. thanks F.~Kahlhoefer  
for useful feedback on ALP astrophysics. L.D. and E.N. acknowledge the PADME collaboration 
for fundamental inputs on the experimental analysis.
L.D. and E.N.~are supported in part by the INFN ``Iniziativa Specifica'' Theoretical 
Astroparticle Physics (TAsP-LNF). M.R. is partly supported by BG-NSF DN-08/14 from 14.12.2016.

\newpage

\medskip

\bibliographystyle{utphys}
\bibliography{ALPleptophi}

\end{document}